%% file: article_SPIE_v6_corrections_hal.tex
\documentclass[]{spie}
\usepackage[svgnames]{xcolor}
\usepackage[greek,english,french]{babel}
\usepackage{pdfpages}
\usepackage{amsmath}
\usepackage{epstopdf}
\usepackage{graphicx}
\usepackage{tikz}
\usetikzlibrary{shapes}
\usetikzlibrary{decorations.pathmorphing}
\graphicspath{{images/}}

\title{Modeling of optical amplifier waveguide based on silicon nanostructures and rare earth ions doped silica matrix gain media by a finite-difference time-domain method: comparison of achievable gain with Er\up{3+} or Nd\up{3+} ions dopants}

\author{Julien Cardin\supit{a}, Alexandre Fafin\supit{a,$\star$}, Christian Dufour\supit{a} and Fabrice Gourbilleau\supit{a}
\skiplinehalf
\supit{a}Centre de Recherche sur les Ions, les Mat\'eriaux et la Photonique (CIMAP), ENSICAEN, UMR 6252 CNRS, CEA/IRAMIS, Universit\'e de Caen Cedex 4,\hspace{1cm} 6 boulevard Mar\'echal Juin, 14050 Caen, France\\
}

\authorinfo{Further author information: (Send correspondence to Julien Cardin)\\Julien Cardin: E-mail: julien.cardin@ensicaen.fr, Telephone: +33(0)2.31.45.26.64\\  $\star$ Present address: Institut P$^{‎\prime}$, D\'epartement Physique et M\'ecanique des Mat\'eriaux, UPR 3346 CNRS, Universit\'e de Poitiers, B\^at. SP2MI, Boulevard Marie et Pierre Curie, BP 30179, 86962 Futuroscope Chasseneuil, France}

\begin{document}
\maketitle

\begin{abstract}
A comparative study of the gain achievement is performed in a waveguide optical amplifier whose active layer is constituted by a silica matrix containing silicon nanograins acting as sensitizer of either neodymium ions (Nd\up{3+}) or erbium ions (Er\up{3+}). Due to the large difference between population levels characteristic times (ms) and finite-difference time step (10$^{-17}$s), the conventional auxiliary differential equation and finite-difference time-domain (ADE-FDTD) method is not appropriate to treat such systems. Consequently, a new two loops algorithm based on ADE-FDTD method is presented in order to model this waveguide optical amplifier. We investigate the steady states regime of both rare earth ions and silicon nanograins levels populations as well as the electromagnetic field for different pumping powers ranging from 1 to $10^4$ mW.mm\up{-2}. Furthermore, the three dimensional distribution of achievable gain per unit length has been estimated in this pumping range. The Nd\up{3+} doped waveguide shows a higher gross gain per unit length at 1064 nm (up to 30 dB.cm\up{-1}) than the one with Er\up{3+} doped active layer at 1532 nm (up to 2 dB.cm\up{-1}). Considering the experimental background losses found on those waveguides we demonstrate that a significant positive net gain can only be achieved with the Nd\up{3+} doped waveguide. The developed algorithm is stable and applicable to optical gain materials with emitters having a wide range of characteristic lifetimes. 
\end{abstract}



\section{INTRODUCTION}
\label{sec:intro}  

For many years, rare earth ions have been used in silica-based optical amplifiers. In these systems, the low gain value requires to employ significant length (10 to 15 m) of doped fiber to achieve a workable power operation. In more compact systems such as erbium-doped waveguide amplifiers (EDWA) a higher gain has to be reached in order to shorten the operating length of the amplifier \cite{Polman2004}. Mostly trivalent erbium ions (Er\up{3+}) have been studied due to their emission wavelength at 1532 nm, which is adapted to the telecommunications window in optical fibers \cite{Miniscalco1991}. However, there are three major gain limiting factors for the erbium ions: the up-conversion mechanism, the excited state absorption and the re-absorption of the signal from the fundamental level. This last drawback is characteristic of a three levels system. More recently, neodymium ions (Nd\up{3+}) has been proposed instead of erbium ion since its four levels  emission scheme prevent signal re-absorption from the fundamental level and makes it more suitable for achieving higher gain.

One main drawback of seeking gain with rare earth ions is their low absorption cross section ($\sigma_{abs}$). However this can be overcome by the use of sensitizers that are characterized by a larger absorption cross section and an efficient energy transfer to RE$^{3+}$ ions. Several sensitizers of RE have been proposed in literature, Polman et al \cite{Polman2004} show the sensitization of Er\up{3+} ions by different kinds of sensitizers such as ytterbium ions, metal ions and silicon nanograins (Si-ng). Several studies\cite{Kenyon1994,Fujii1997} have pointed out that silicon nanograins are efficient sensitizers of Er\up{3+} ions and can increase its effective absorption cross section by a factor up to $10^4$. MacDonald et al\cite{macdonald2006luminescence} showed a likewise efficient sensitization of Nd\up{3+} ions by those Si-ng sensitizers.  
 
We aim to model the propagation of an electromagnetic field into a waveguide with a layer containing absorbing and emitting centers as for example rare earth ions and silicon nanograins. More particularly, we want to determine the system characteristics as fields, level populations, and gain in a steady state regime as a function of initial parameters such as concentration of emitting centers, geometry, pumping configuration and pump and signal powers. Moreover, to model those steady states regimes in a waveguide with a layer containing absorbing and emitting centers, we must take into account the time evolution of electromagnetic field and electronic levels populations of these centers. In a waveguide containing silicon nanograins and rare earth ions, the typical lifetime of the electronic levels is about some ms, whereas the characteristic period of the electromagnetic field is of the order of fs. The steady state of the system cannot be reached with such a difference between absorbing and emitting centers lifetimes and electromagnetic field period by means of classical ADE-FDTD. It would require prohibitively long computation times with about 10$^{15}$ iterations. Consequently, we developed a two loops new algorithm based on ADE-FDTD method which allows to compute the electromagnetic field distribution, the electronic levels populations, and the gain in the waveguide in the steady state regime \cite{fafin2013modeling,fafin2014theoretical}. In this paper we present briefly the two loops ADE-FDTD algorithm of calculation and the comparative study of waveguides with an active layer containing Si-ng and doped either with erbium or neodymium ions.

\section{Rare earth doped silicon based waveguide}
\begin{figure}[htbp]
\centering
\includegraphics[scale=0.78]{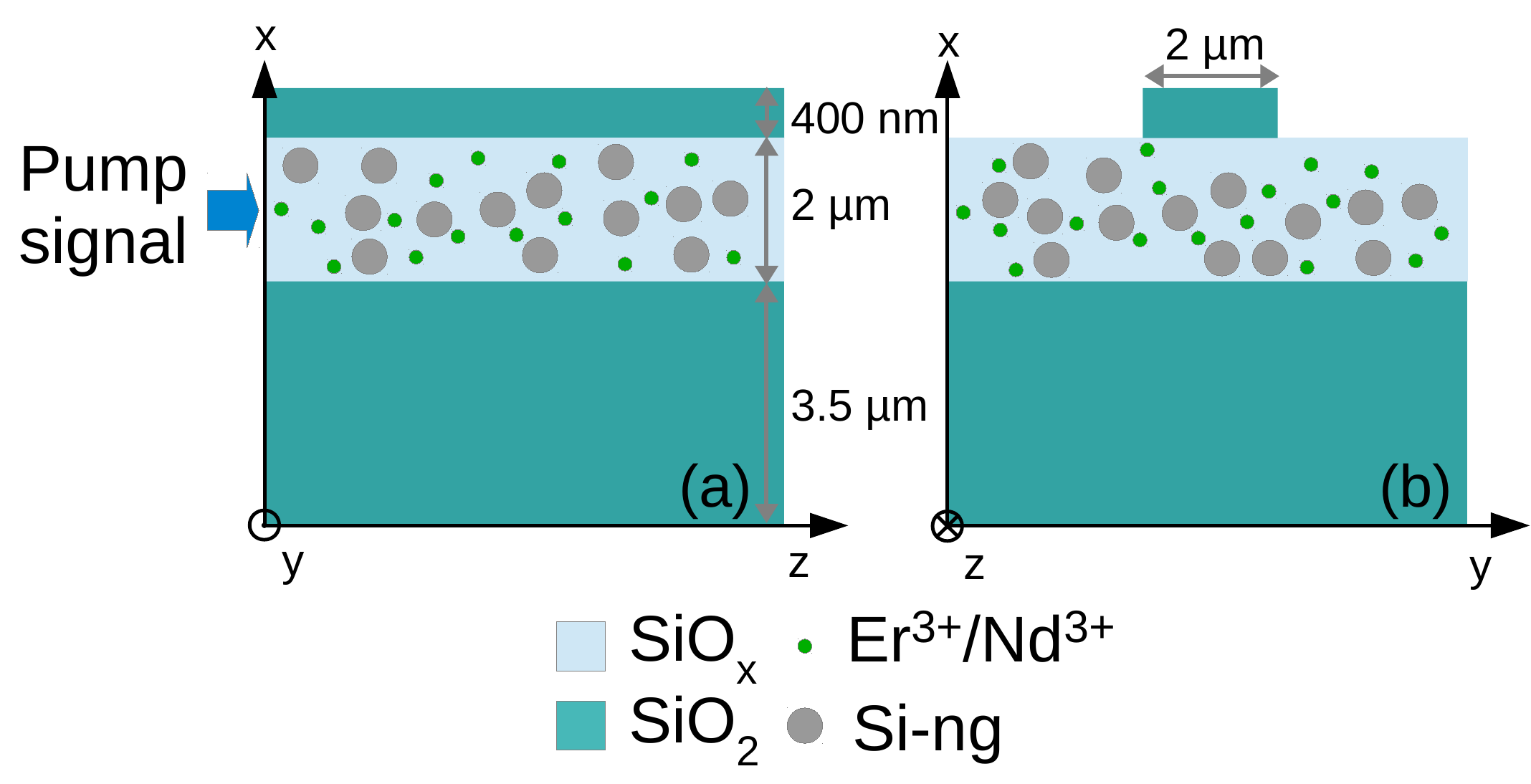}
\caption{Optical amplifier waveguide based silicon rich silicon oxide doped with rare earth ions : (a)in longitudinal view  with configuration of co-propagation of pump and signal and (b) in cross section view}
\label{waveguide}
\end{figure}

The system describe hereafter is an optical amplifier strip loaded waveguide, composed of three layers as presented on figure \ref{waveguide}. Bottom and top cladding layers of pure silica of a thickness equal to 3.5 $\mu$m and 0.4 $\mu$m respectively ensure vertical guiding of modes, whereas a top silica strip stacked on the active layer ensures lateral confinement of modes. The static refractive index (i.e. refractive index which remains constant with wavelength) of the active layer is equal to 1.5 and the one of the strip and bottom cladding layers to 1.448 to ensure the guiding conditions. We inject at one end of the waveguide pump and signal guide modes in a co-propagation scheme as presented on figure 1(a). The active layer of 2 $\mu$m thick is constituted of non-stoichiometric Silicon oxide (SiO$_x$) containing silicon nanograins and rare earth ions (RE\up{3+}).

\begin{figure}[htbp]
   \begin{minipage}[c]{.5\linewidth}
   \center
   \scalebox{1.1}{\input{SRSO_Er2.tex}}
   \end{minipage} \hfill
   \begin{minipage}[c]{.5\linewidth}
   \center
   \scalebox{1.1}{\input{SRSO_Nd2.tex}}
   \end{minipage}
   \caption{Schematic representation of the principle of optical indirect excitation of the rare earth ions via sensitization by silicon nanograin in silicon rich silicon oxide sample with silicon nanograins and doped with : (a) erbium ion (b) neodymium ion}
   \label{SRSO:RE}
\end{figure}
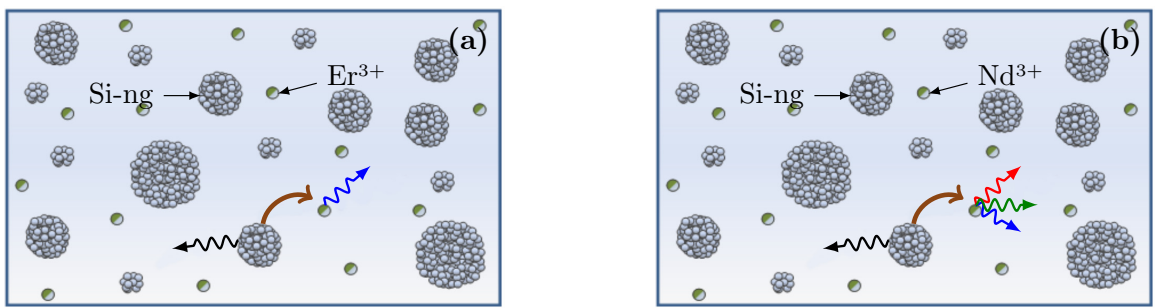

The active layers doped with RE\up{3+} are grown by magnetron sputtering technique as described in references \cite{hijazi2009structural,cueff2009impact,liang2013evidence}. Those composite layers contain silicon nanograins (Si-ng) and RE\up{3+} ions dispersed in non-stoichiometric Silicon oxide (SiO$_x$) as schematically presented of figure \ref{SRSO:RE}. Those Si-ng act as sensitizers of RE\up{3+} ions in their vicinity by a mechanism of energy transfer and therefore enhance their excitability by several order of magnitude. The effective absorption cross section of RE\up{3+} sensitized by Si-ng is of the order of $10^{-16}$ cm$^2$ against RE\up{3+} direct excitation absorption cross section which is in the  $10^{-19}$ to $10^{-21}$ cm$^2$ range. According to our experimental investigations \cite{Debieu2011} an exciton is photogenerated in Si-ng by pumping of SiO$_x$:Si-ng:RE\up{3+} layer with an electromagnetic wave at 488 nm. After non-radiative deexcitations, the exciton energy reaches the corresponding energy gap of RE\up{3+}. Exciton can then vanish either by transfer of energy to the RE\up{3+} ions by dipole-dipole interaction or by radiative or non-radiative recombination to Si-ng ground level.

\begin{figure}[htbp]
   \begin{minipage}[c]{.5\linewidth}
   \center
   \includegraphics[scale=0.325]{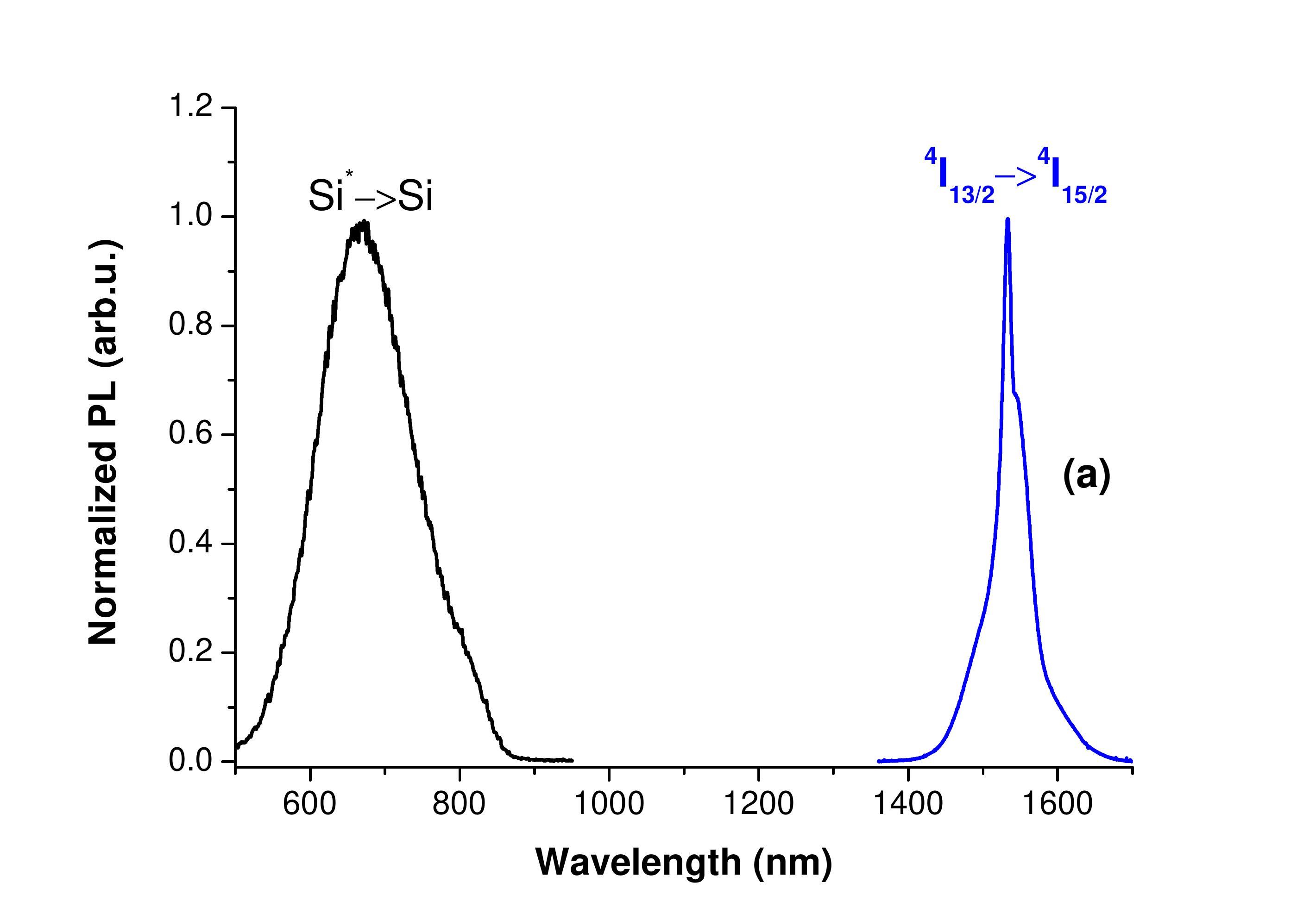}
   \end{minipage} \hfill
   \begin{minipage}[c]{.5\linewidth}
   \center
   \includegraphics[scale=0.325]{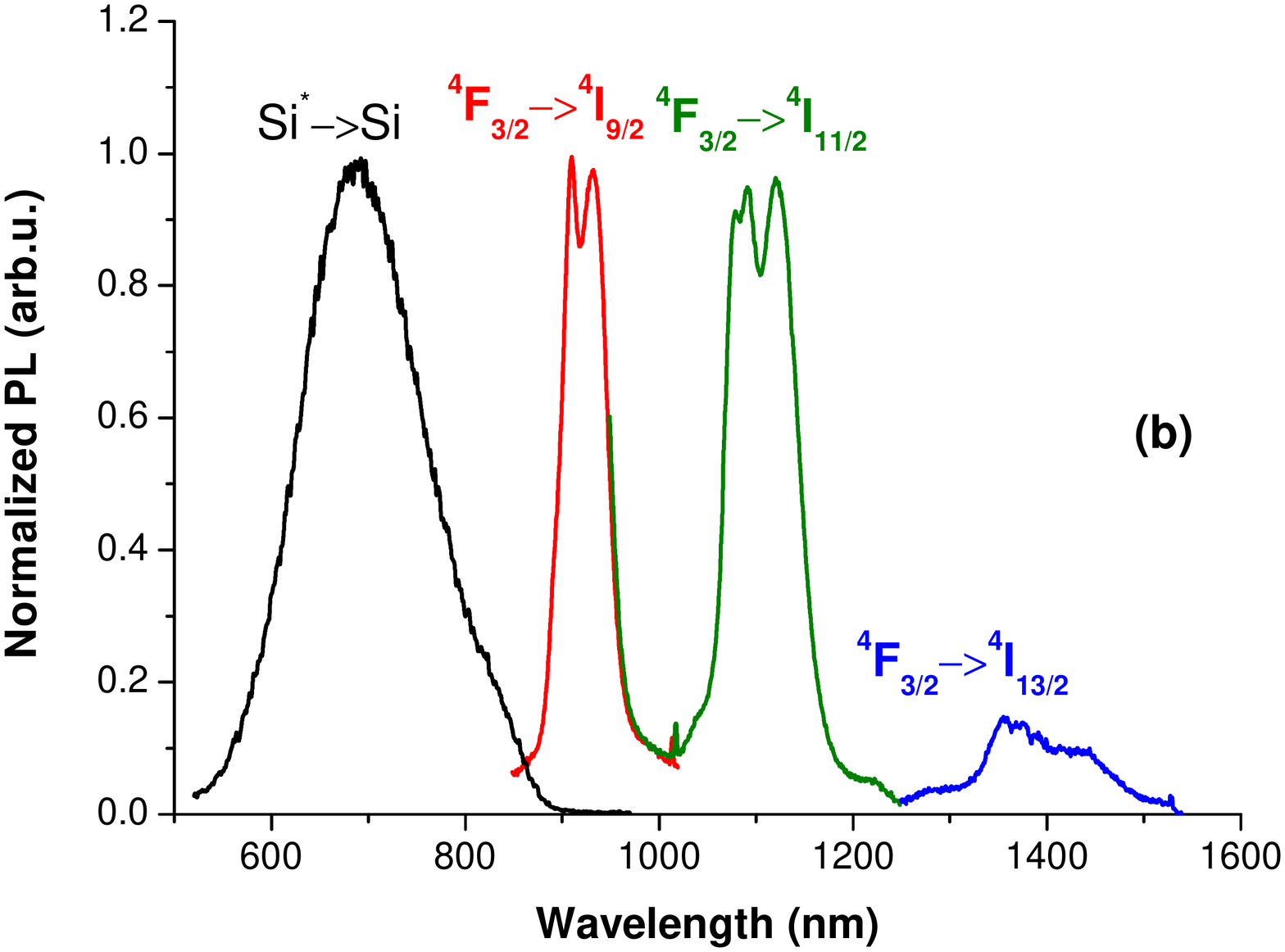}
   \end{minipage}
   \caption{Typical photoluminescence spectra of silicon rich silicon oxide sample containing silicon nanograins and doped with rare earth ions: (a) erbium ions (b) neodymium ions}
    \label{PLSRSO:RE}
\end{figure}

Hence those active layers have typical/particular photoluminescence properties as presented on figure \ref{PLSRSO:RE}. The characteristic emission band of radiative transition occurring from excited (Si\up{*}) to ground state (Si) in Si-ng is at 770nm. The characteristic emission bands of Er\up{3+} and Nd\up{3+} ions are respectively at  1532 nm and at 921, 1106 and 1395 nm.

\section{Two loops ADE-FDTD method}

The FDTD method is based on time and space discretization scheme of Maxwell equations proposed by Yee \cite{Yee1966} which allows to calculate the propagation of electromagnetic field (\textbf{E},\textbf{H}) in time domain \cite{Taflove1995}. The ADE method consists in the use of extra terms such as current density \textbf{J} or polarization density \textbf{P} which are solutions of differential equations with the aim to model some non linear optical behavior such as dispersive or gain media \cite{Nagra1998,hagness1996subpicosecond}. The Maxwell equations describing fields \textbf{E}, \textbf{H} are rewritten in the form given in equation \ref{equations maxwell}.
\begin{equation}
		\left\{
		\begin{split}
		\nabla \wedge \textbf{E}&= -\mu_0\mu_r \frac{\partial \textbf{H}}{\partial t} - \rho\textbf{H} \\
		\nabla \wedge \textbf{H}&= \epsilon_0\epsilon_r\frac{\partial \textbf{E}}{\partial t} + \frac{\partial \textbf{P}_{tot}}{\partial t}+ \sigma \textbf{E}
		\end{split}
		\right.	 
		\label{equations maxwell}
\end{equation}
In equation \ref{equations maxwell}, $\epsilon_r$ and $\mu_r$ are respectively the static dielectric permittivity and magnetic permeability of different considered materials. $\sigma$ is the electrical conductivity and $\rho$ is a fictitious magnetic resistivity used for boundary conditions of the calculation box. The perflectly matched layers (PML) scheme developped by Berenger\cite{Berenger1994} have been implemented as boundary conditions. $\mathbf{P}_{tot} = \sum \mathbf{P}_{ij}$ is the sum of all polarizations corresponding to each transition of absorbing/emitting centers (hereafter: silicon nanograins and rare earth ions). The use of one polarization density $\mathbf{P}_{ij}$ per optical transition between levels i and j allows the description of the active layer dynamic permittivity $\epsilon(\omega)$ arising from the dipole moment densities induced by optical transitions in emitting centers. The FDTD space step $\Delta$, time step $\Delta_t$, Courant number $S_c$ and calculation box size are reported in Table \ref{FDTDparam}.
\begin{table}[htbp]
	\footnotesize
	\centering
	\caption{FDTD algorithm parameters}
	\begin{tabular}{|c|c|c|c|c|c|c|}
	\hline
	Parameter & $\Delta$ & $\Delta_t$ & $S_c$ & box length x& box length y& box length z\\
	\hline	
	Value   &$50$ nm & $10^{-17}$ s& $1$ & $169\Delta$ &$321\Delta$ &$161\Delta$\\
	\hline
	\end{tabular}
	\label{FDTDparam}
\end{table}

\begin{figure}[htbp]
   \begin{minipage}[c]{.35\linewidth}
   \scalebox{0.95}{\input{algoclassique.tex}}
   \end{minipage} \hfill
   \begin{minipage}[c]{.65\linewidth}
   \scalebox{0.95}{\input{nouvelalgo.tex}}
   \end{minipage}
   \caption{Flowcharts showing (a) the classical ADE-FDTD algorithm and (b) the new hybrid-algorithm with the alternation of the short times loop calculating electromagnetic field and polarizations and the long times loop including calculation of levels populations.}
   \label{algo}
\end{figure}
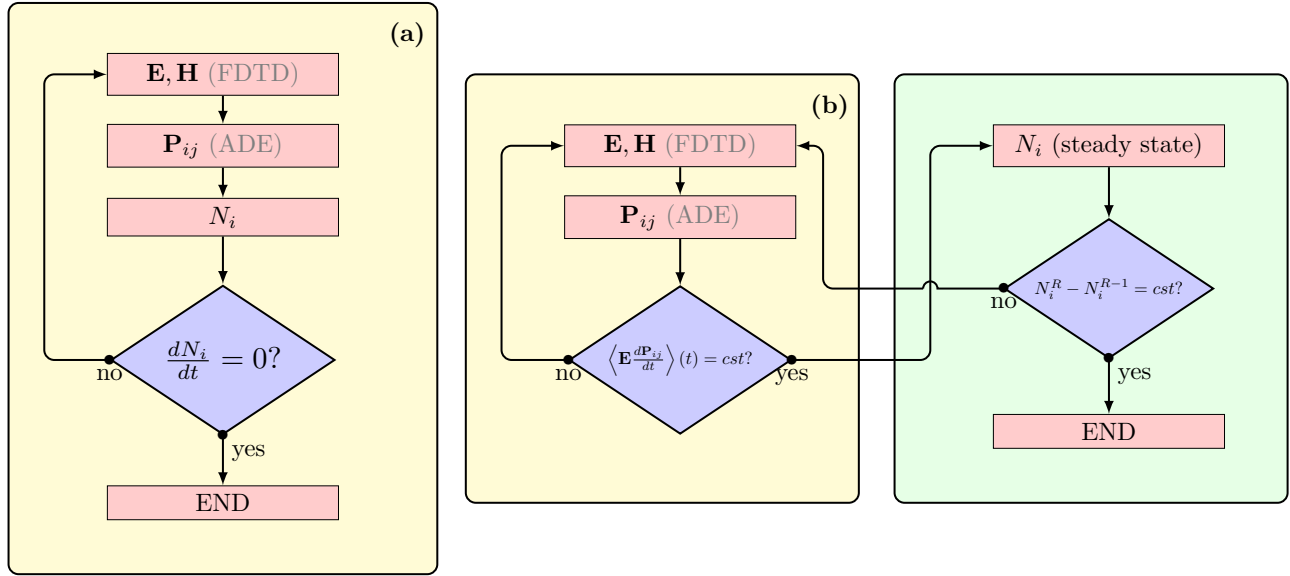

Neglecting the Rabi oscillation term \cite{chang2004finite}, for a transition between levels i and j the polarization density $\mathbf{P}_{ij}$ is linked to the instantaneous electric field \textbf{E}(t), to the population difference $\Delta N_{ij}=N_i-N_j$, and to a constant $\kappa_{ij}=\frac{6 \pi \epsilon_0 c^3}{\omega_{ij}^2 \tau_{ij} n}$ through the Lorentz type polarization density differential equation \cite{Siegman1986}:
\begin{equation}
\frac{d^{2} \textbf{P}_{ij}(t)}{dt^2}+\Delta \omega_{ij} \frac{d \textbf{P}_{ij}(t)}{dt}+\omega_{ij}^2 \textbf{P}_{ij}(t) = \kappa_{ij} \Delta N_{ij} (t) \textbf{E}(t)
\label{polarisation}
\end{equation}
The time evolution of electronic levels population N$_i$ is modeled by classical rate equations system. Full detailed rate equations system of considered emitters will be given thereafter. As mentioned in the introduction, some excited levels characteristic lifetimes may be as long as a few ms \cite{Pacifici2003,Lee2005}. Due to the time step of ADE-FDTD classical method (figure \ref{algo}a), which are lower than $10^{-17}$ s \cite{Taflove1995}, the number of iteration should be as huge as $10^{15}$ in order to reach the steady states of the levels populations. A conventional calculation is therefore impossible in a reasonable time. Consequently a new two loops algorithm based on ADE-FDTD method (figure \ref{algo}b) has been developed that overcomes the multi-scale times issue of such a system and allows to describe the spatial distribution of the electromagnetic field and levels population in steady state in an active optical waveguide \cite{fafin2013modeling,fafin2014theoretical}. This algorithm works with a drastic reduction of number of iterations to reach steady states values of fields and levels population from about $10^{15}$ to 10$^5$ iterations.

\begin{figure}[htbp]
\centering
\includegraphics[scale=0.35]{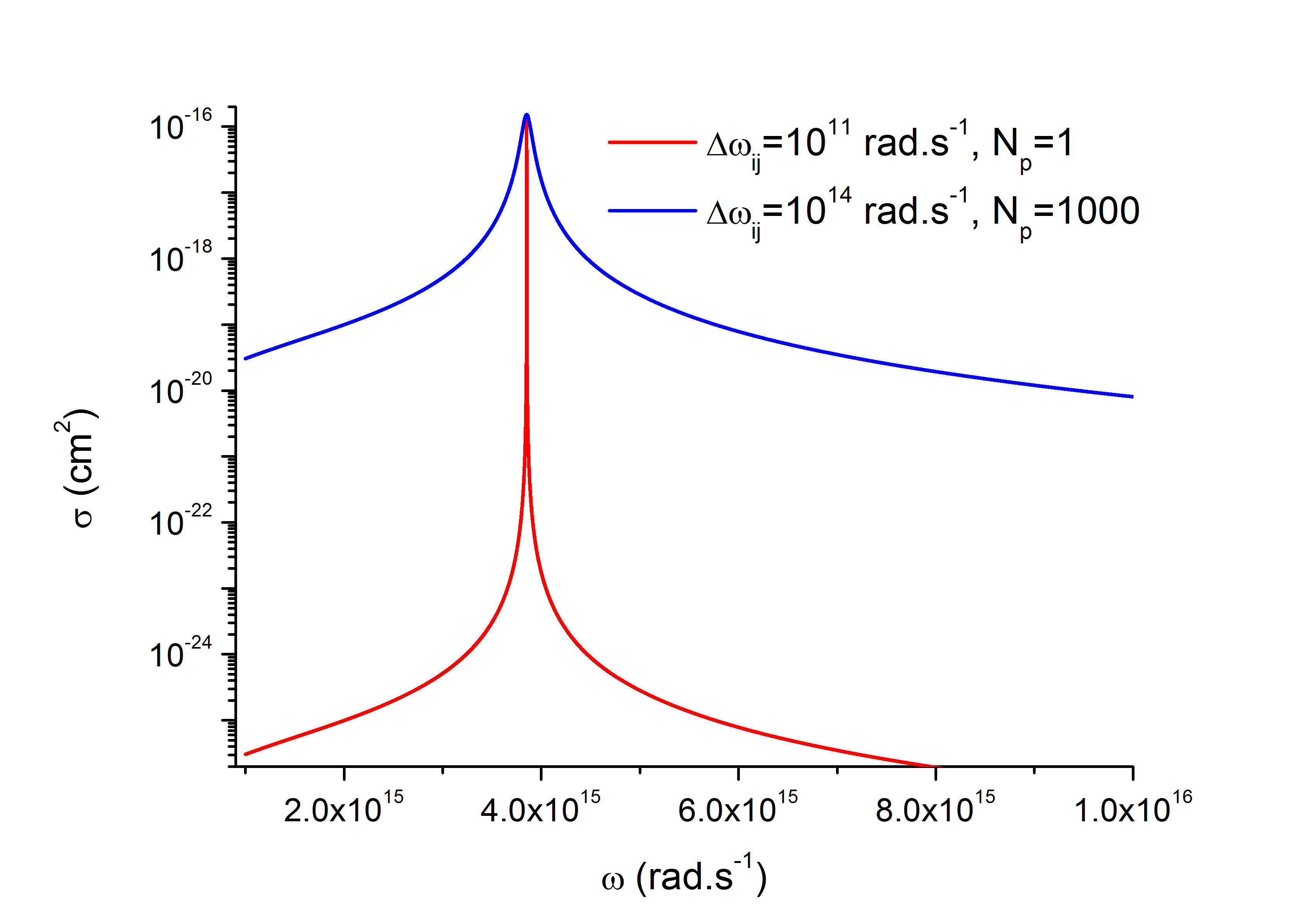}
\caption{Absorption cross section as a function of the pulsation with a transition at $ 3.8 \times 10^{15}~\mathrm{rad.s^{-1}}$ for two number of polarization densities N$_p$ and two linewidths $\Delta\omega_{ij}$}
\label{SuperPolarSC}
\end{figure}

The description by the Lorentz oscillator polarization density (Equation 2) of an absorption or an emission process occurring during a transition $i$ to $j$ implies the equality of absorption and emission cross sections and only homogeneous broadening of the transition linewidth. At the resonance frequency, each Lorentz oscillator linewidth $\Delta\omega_{ij}$ is related to the absorption/emission cross section by the equation \ref{cross section}.
\begin{equation}
\sigma_{ij} =\frac{\kappa_{ij}}{\epsilon_0 c} \frac{1}{\Delta \omega_{ij}}
\label{cross section}
\end{equation}
According to equation \ref{cross section}, high absorption cross sections $\sigma_{ij}$ (typically greater than $10^{-17}$ cm\up{2}) lead to a small linewidth $\Delta\omega_{ij}$ (about $10^{11}$ rad.s\up{-1}) which impose a large number of iterations to reach a steady state. In order to calibrate the proper absorption cross section while keeping the number of iterations as small as possible, we exploit the superposition property of the polarization densities by using a number N$_p$ of identical polarization densities with larger $\Delta\omega_{ij}$. Hence, equation (\ref{cross section}) becomes equation (\ref{cross section2}).
\begin{equation}
\sigma_{ij} = \frac{\kappa_{ij}}{\epsilon_0 c} \frac{N_p}{\Delta \omega_{ij}}
\label{cross section2}
\end{equation}
At resonance frequency for $ \omega_{ij}=3.8 \times 10^{15}~\mathrm{rad.s^{-1}}$, the two methods lead to identical cross sections : $\sigma_{ij}(N_p=1)=\sigma_{ij}(N_p=1000)$ as represented on figure \ref{SuperPolarSC}. 
By this method, the cross section calibration is performed by choosing the appropriate value of the number of identical polarization densities $N_p$ and the linewidth $\Delta \omega_{ij}$.

\begin{figure}[htbp]
   \begin{minipage}[c]{.5\linewidth}
   \scalebox{0.89}{\input{schema_erbium_complet.tex}}
   \end{minipage} \hfill
   \begin{minipage}[c]{.5\linewidth}
   \scalebox{0.89}{\input{schema_neodyme_complet.tex}}
   \end{minipage}
   \caption{Excitation mechanism of (a) erbium ions and (b) neodymium ions in silicon rich silicon oxide containing silicon nanograins and rare earth ions. Radiative transitions are represented by straight lines and nonradiative ones by sinusoidal lines.}
   \label{bandes}
\end{figure}
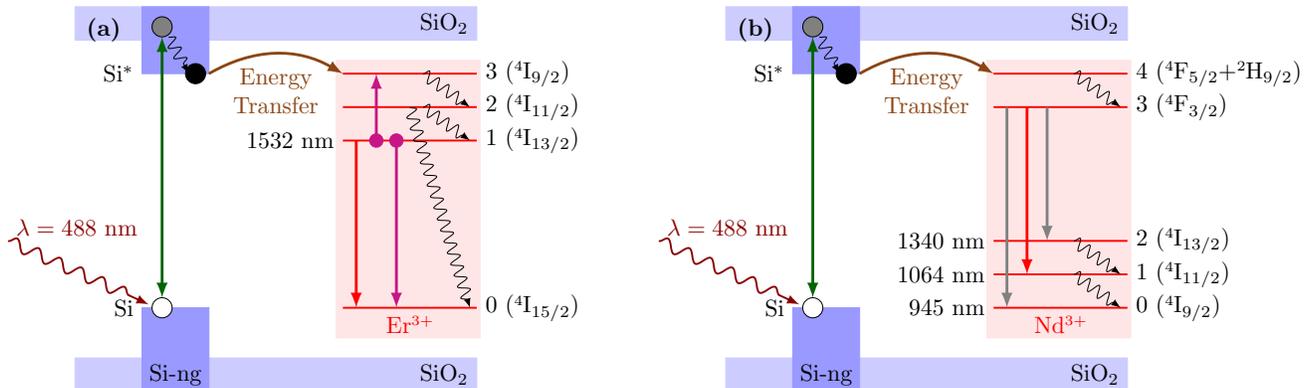

Silicon nanograins (Si-ng) are modeled as a two levels system (Figure \ref{bandes}) where the ground and excited levels populations (respectively N$_{Si}$ and N$_{Si^*}$) are given by the rate equations (Equation \ref{NSi1} and \ref{NSi0}). Due to a low probability of multi-exciton generation  in a single Si-ng \cite{Govoni2012}, we assume the excitation of one single exciton by Si-ng, therefore the Si-ng population will correspond to the exciton population. 

\begin{equation}
\dfrac{dN_{Si^*}\left(t\right)}{dt}=+\frac{1}{\hbar\omega_{Si}}\textbf{E}\left(t\right)\frac{d\textbf{P}_{Si} \left(t\right)}{dt}-\dfrac{N_{Si^*}\left(t\right)}{\tau_{Si}\vert _{nr}^{r}}-KN_{Si^*}\left(t\right)N_{0}\left(t\right)
\label{NSi1}
\end{equation}

\begin{equation}
\dfrac{dN_{Si}\left(t\right)}{dt}=-\frac{1}{\hbar\omega_{Si}} \textbf{E}\left(t\right)\frac{d\textbf{P}_{Si}\left(t\right)}{dt}+\dfrac{N_{Si^*}\left(t\right)}{\tau_{Si}\vert _{nr}^{r}}+KN_{Si^*}\left(t\right)N_{0}\left(t\right)
\label{NSi0}
\end{equation}
The Si-ng physical parameters are reported in table 2 (for further detail see references \cite{fafin2013modeling} and \cite{fafin2014theoretical}).
\begin{table}[htbp]
	\footnotesize
	\caption{Parameters levels of silicon nanograins}
	\centering
	\begin{tabular}{|c|c|c|c|c|c|c|}
		\hline
		$j \rightarrow i$ & Lifetime (s) & $ \omega_{ij} (10^{15}~\mathrm{rad/s}) $ & $ \Delta \omega_{ij} (10^{14}~\mathrm{rad/s}) $ & $ N_p $&  K ($cm^{-3}.s^{-1}$)&[Si-ng] ($cm^{-3}$)\\
		\hline
		$ Si^*\rightarrow Si $ & $ 50.10^{-6} $ & 3.682 & 1 & 2756 &$10^{14}$&$10^{19}$\\
		\hline
	\end{tabular}
	\label{silicon parameters}
\end{table}

Erbium ions levels diagram is presented on figure \ref{bandes}a. We consider two non-radiative transitions and one radiative transition from level 1 to level 0 at 1532 nm. Moreover an up-conversion process from level 1 to levels 0 and 3 can be modeled. The time evolution of Er\up{3+} levels populations is described by the rate equations \ref{eq_er_3} to \ref{eq_er_0}. The concentration of erbium ions was taken equal to $10^{20}$ cm\up{-3}. All parameters of erbium ions transitions are taken from references\cite{Pacifici2003,Toccafondo2008} and reported in Table \ref{erbium parameters}.
\begin{align}
\label{eq_er_3}
\dfrac{dN_{3}\left(t\right)}{dt}=&-\dfrac{N_{3}\left(t\right)}{\tau_{32}\vert_{nr}}+KN_{Si^*}\left(t\right)N_{0}\left(t\right)+C_{up} N_1^2 \\
\label{eq_er_2}
\dfrac{dN_{2}\left(t\right)}{dt}=&+\dfrac{N_{3}\left(t\right)}{\tau_{32}\vert_{nr}}-\dfrac{N_{2}\left(t\right)}{\tau_{21}\vert_{nr}}-\dfrac{N_{2}\left(t\right)}{\tau_{20}\vert_{nr}} \\
\label{eq_er_1}
\dfrac{dN_{1}\left(t\right)}{dt}=&+\frac{1}{\hbar\omega_{10}}\textbf{E}\left(t\right)\frac{d\textbf{P}_{10}\left(t\right)}{dt}+\dfrac{N_{2}\left(t\right)}{\tau_{21}\vert_{nr}}-\dfrac{N_{1}\left(t\right)}{\tau_{10}\vert_{nr}^r}-2 C_{up} N_1^2 \\
\label{eq_er_0}
\dfrac{dN_{0}\left(t\right)}{dt}=& -\frac{1}{\hbar\omega_{10}}\textbf{E}\left(t\right)\frac{d\textbf{P}_{10}\left(t\right)}{dt} \\ \nonumber &+\dfrac{N_{2}\left(t\right)}{\tau_{20}\vert_{nr}}+\dfrac{N_{1}\left(t\right)}{\tau_{10}\vert_{nr}^r}-KN_{Si^*}\left(t\right)N_{0}\left(t\right)+C_{up} N_1^2
\end{align}

\begin{table}[htbp]
	\footnotesize
	\caption{Erbium ions parameters}
	\centering
	\begin{tabular}{|c|c|c|c|c|}
	\hline
$j \rightarrow i$  & 3$\rightarrow$2   & 2$\rightarrow$1     & 2$\rightarrow$0     & 1$\rightarrow$0 \\
\hline
Lifetime (s)& $0.1\times10^{-6}$   & $2.4\times10^{-6}$& $ 710\times10^{-6} $& $ 8.5\times10^{-3} $\\\hline
$\omega_{ij}$ ($10^{15}~\mathrm{rad/s}$) &             &   1.23              &                     &   \\\hline
$\Delta \omega_{ij}$ ($10^{15}~\mathrm{rad/s}$)&       &   0.15              &                     &    \\\hline
$ N_p $	&       &   1              &                     &  \\\hline
$ C_{up} $	($cm^{3}.s^{-1}$)&       &                 &                     &  5$\times$10$^{-17}$\\\hline
	\end{tabular}
	\label{erbium parameters}
\end{table}

Neodymium ions levels diagram is presented on figure \ref{bandes}b. We consider three non-radiative transitions and three radiative transitions and neglect up-conversion phenomena. The concentration of neodymium ions is taken equal to $10^{20}$ cm\up{-3}. The time evolution of Nd\up{3+} levels populations is described by the rate equations \ref{eq_nd_4} to \ref{eq_nd_0}. Parameters of neodymium ions transitions are taken from \cite{serqueira2006judd,Siegman1986} and reported in Table \ref{neodymium parameters}.

\begin{align}
\label{eq_nd_4}
\frac{dN_{4}\left(t\right)}{dt}=&-\frac{N_{4}\left(t\right)}{\tau_{43}\vert_{nr}}+KN_{Si^*}\left(t\right)N_{0}\left(t\right) \\
\label{eq_nd_3}
\frac{dN_{3}\left(t\right)}{dt}=&+\frac{1}{\hbar\omega_{30}}\textbf{E}\left(t\right)\frac{d\textbf{P}_{30}\left(t\right)}{dt}+\frac{1}{\hbar\omega_{31}}\textbf{E}\left(t\right)\frac{d\textbf{P}_{31}\left(t\right)}{dt}+\frac{1}{\hbar\omega_{32}}\textbf{E}\left(t\right)\frac{d\textbf{P}_{32}\left(t\right)}{dt} \\ \nonumber &+\frac{N_{4}\left(t\right)}{\tau_{43}\vert_{nr}}-\frac{N_{3}\left(t\right)}{\tau_{30}\vert_{nr}^r}-\frac{N_{3}\left(t\right)}{\tau_{31}\vert_{nr}^r}-\frac{N_{3}\left(t\right)}{\tau_{32}\vert_{nr}^r} \\
\label{eq_nd_2}
\frac{dN_{2}\left(t\right)}{dt}=&-\frac{1}{\hbar\omega_{32}}\textbf{E}\left(t\right)\frac{d\textbf{P}_{32}\left(t\right)}{dt}+\frac{N_{3}\left(t\right)}{\tau_{32}\vert_{nr}^r}-\frac{N_{2}\left(t\right)}{\tau_{21}\vert_{nr}} \\
\label{eq_nd_1}
\frac{dN_{1}\left(t\right)}{dt}=&-\frac{1}{\hbar\omega_{31}}\textbf{E}\left(t\right)\frac{d\textbf{P}_{31}\left(t\right)}{dt}+\frac{N_{3}\left(t\right)}{\tau_{31}\vert_{nr}^r}-\frac{N_{1}\left(t\right)}{\tau_{10}\vert_{nr}}+\frac{N_{2}\left(t\right)}{\tau_{21}\vert_{nr}} \\
\label{eq_nd_0}
\frac{dN_{0}\left(t\right)}{dt}=&-\frac{1}{\hbar\omega_{30}}\textbf{E}\left(t\right)\frac{d\textbf{P}_{30}\left(t\right)}{dt}+\frac{N_{3}\left(t\right)}{\tau_{30}\vert_{nr}^r}+\frac{N_{1}\left(t\right)}{\tau_{10}\vert_{nr}}-KN_{Si^*}\left(t\right)N_{0}\left(t\right)
\end{align}

\begin{table}[htbp]
	\footnotesize
	\caption{Neodymium ion parameters}
	\centering
	\begin{tabular}{|c|c|c|c|c|c|c|}
		\hline
		$ j \rightarrow i$  & 4$\rightarrow$3 & 3$\rightarrow$2 & 3$\rightarrow$1 & 3$\rightarrow$0 & 2$\rightarrow$1 & 1$\rightarrow$0 \\
		\hline
		Lifetime (s)& $230\times10^{-12}$  & $ 1000\times10^{-6} $   & $200\times10^{-6}$   & $250\times10^{-6}$ & $970\times10^{-12}$ & $510\times10^{-12}$  \\\hline
		$\omega_{ij}$ ($10^{15}~\mathrm{rad/s}$) &   &   1.34   & 1.77 & 1.99 &  &  \\\hline
		$\Delta \omega_{ij}$ ($10^{15}~\mathrm{rad/s}$)&  & 0.67& 0.18 & 0.11&  &  \\\hline
		$ N_p $ &  & 1& 1 & 1&  &\\
		\hline
	\end{tabular}	
	\label{neodymium parameters}
\end{table}

We aim at calculating levels populations in steady states in a faster and more convenient manner. Therefore, the preferred method would be to analytically calculate the steady state populations by setting $\frac{dN_i}{dt}=0$ in each equation. Consequently, this method is applied in case of neodymium levels equations system. In erbium rate equations system a finite-difference method was used in order to reach the steady states of population levels. This calculation was performed using a time step ten times lower than the shortest lifetime (0.1 $\mu$s) of the rate equations system. The calculation time is then no longer negligible but does not rise up significantly the global calculation time of the two loops ADE-FDTD method.

\section{Results and discussion}

\begin{figure}[htbp]
   \begin{minipage}[c]{.5\linewidth}
   \center
   \includegraphics[scale=0.675]{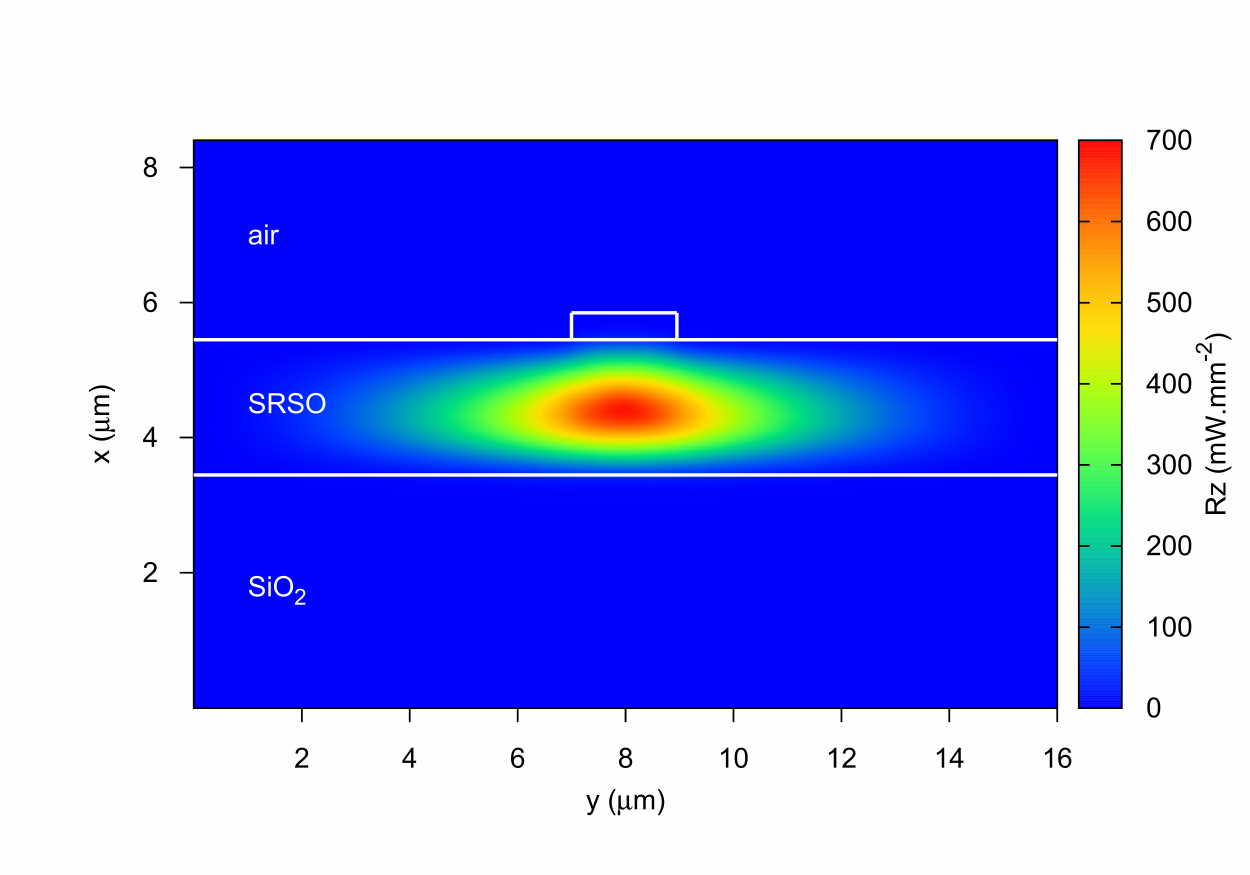}
   \end{minipage} \hfill
   \begin{minipage}[c]{.5\linewidth}
   \center
   \includegraphics[scale=0.675]{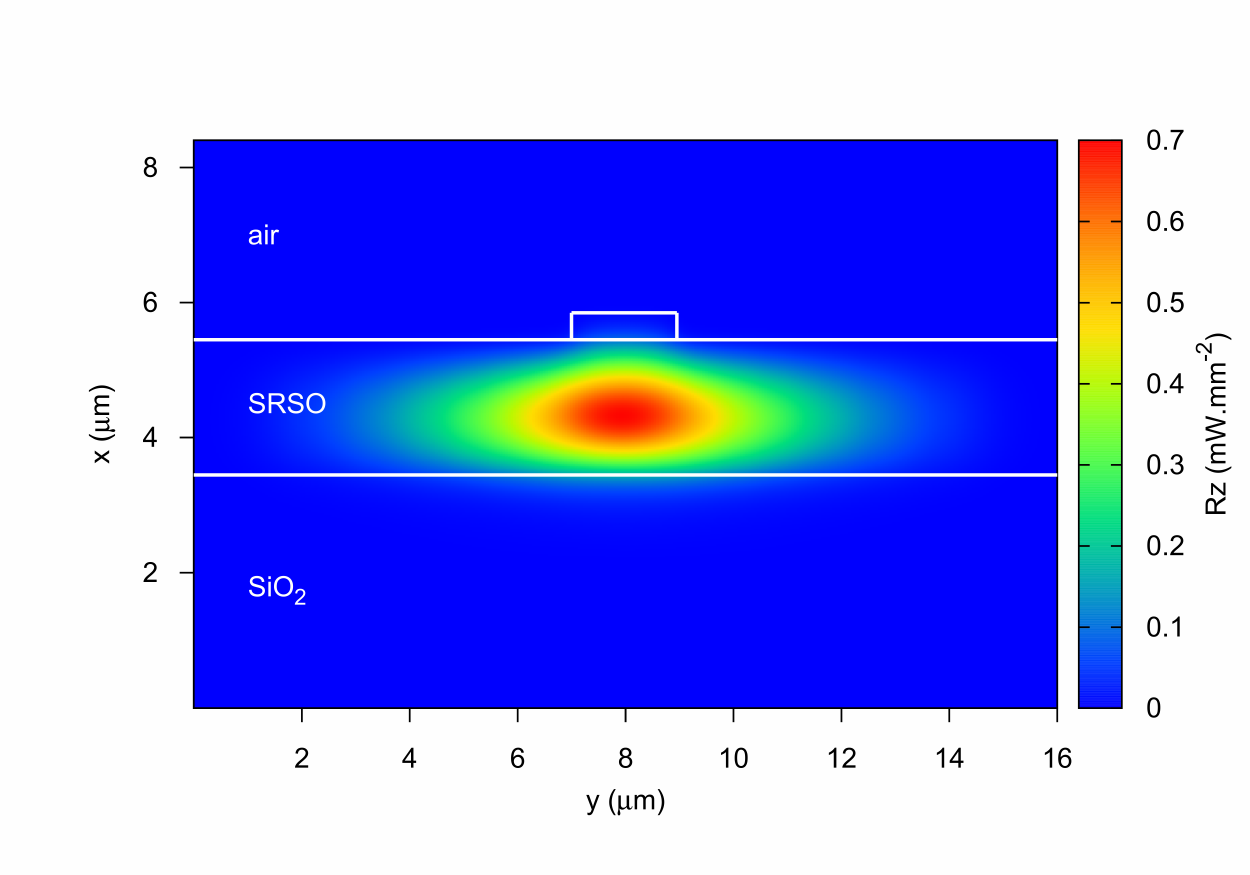}
   \end{minipage}
   \caption{Transverse section views of Z-component of the time average Poynting vector of the pump (on the left) and of the signal (on the right) in the middle of the waveguide. The pump  wavelength is $\lambda_p$=488 nm and the signal $\lambda_s$=1064 nm. Injected powers are respectively 1 $W.mm^{-2}$ and 1 mW.mm\up{-2}}
\label{trans}
\end{figure}

\begin{figure}[htbp]
   \begin{minipage}[c]{.5\linewidth}
   \center
   \includegraphics[scale=0.675]{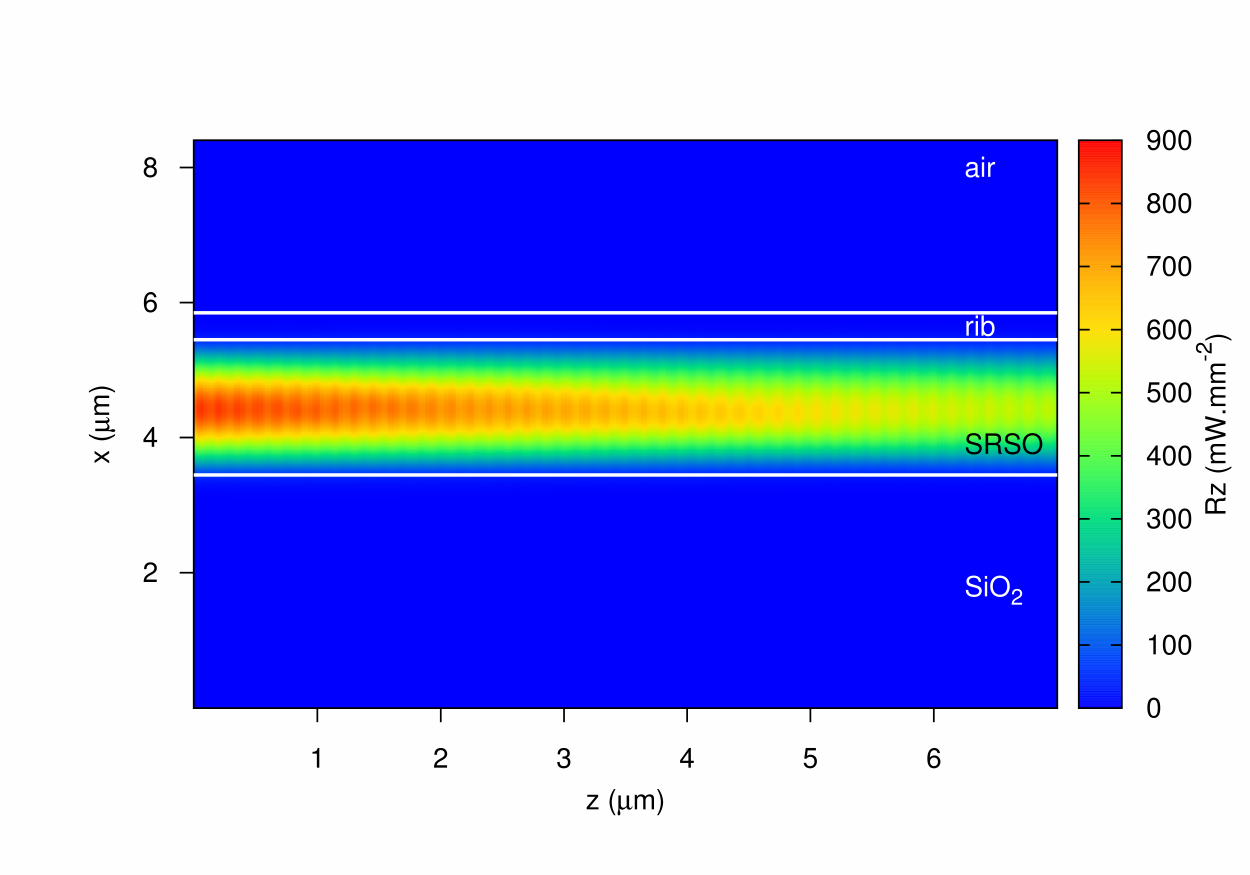}
   \end{minipage} \hfill
   \begin{minipage}[c]{.5\linewidth}
   \center
   \includegraphics[scale=0.675]{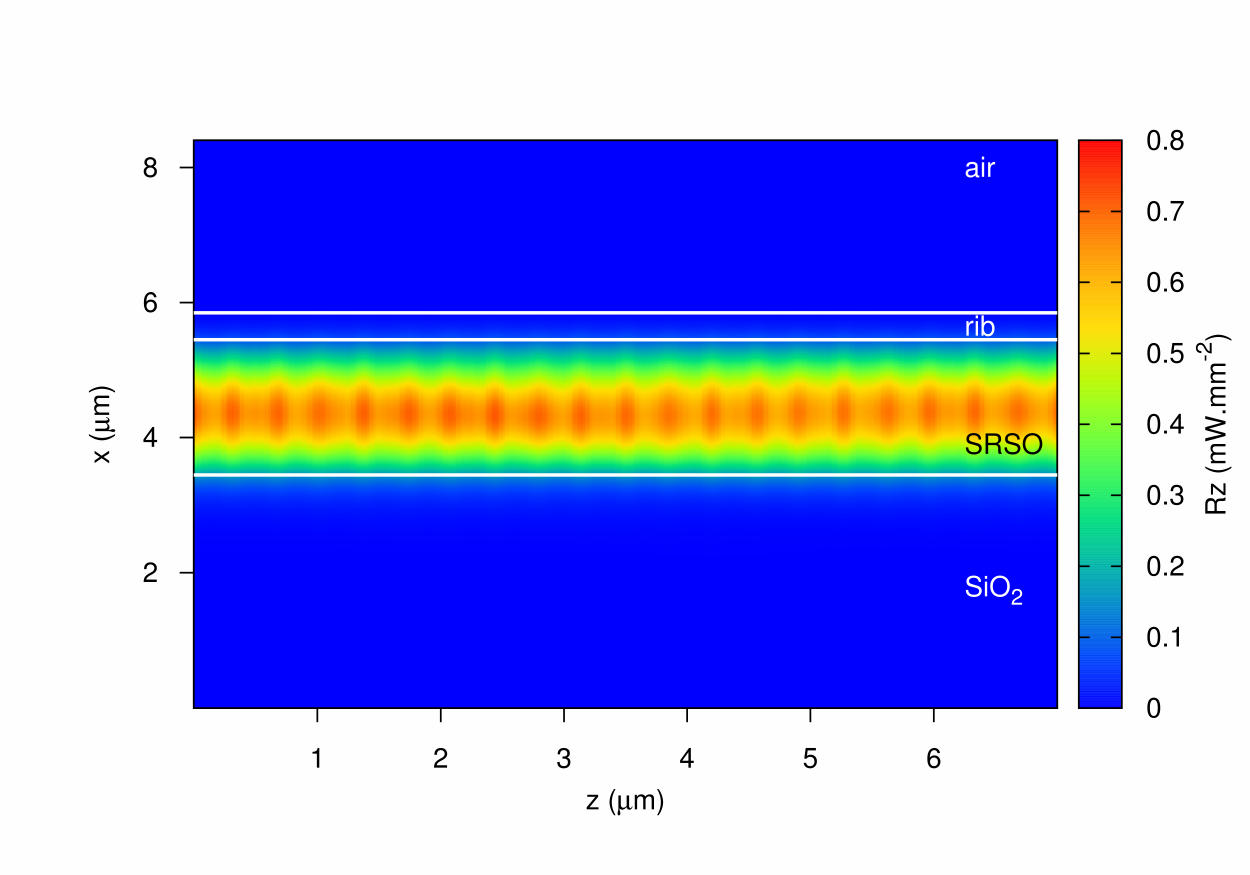}
   \end{minipage}
   \caption{Longitudinal section views of Z-component of the time average Poynting vector of the pump (on the left) and of the signal (on the right) in the middle of the waveguide. The pump wavelength is $\lambda_p$=488 nm and the signal $\lambda_s$=1064 nm. Injected powers are respectively 1 $W.mm^{-2}$ and 1 mW.mm\up{-2}}
\label{longi}
\end{figure}

Poynting vector ($\mathbf{R} = \mathbf{E}\times \mathbf{H}$) has been calculated from time Fourier transform of electromagnetic field (\textbf{E}, \textbf{H}) computed by ADE-FDTD method. Both signal and pump were injected as guided mode in the waveguide. We determine the main component \textbf{R}$_\mathrm{z}$ of the guided pump (488 nm) and signal (1064 nm) power density of $\mathbf{R}$ in transverse and longitudinal section view (in W.mm\up{-2}) on figures \ref{trans} and \ref{longi} respectively. A decrease of the pump \textbf{R}$_\mathrm{z}^{pump}$ along the waveguide is observed, which is due to the absorption at 488 nm by the silicon nanograins. Moreover, the signal \textbf{R}$_\mathrm{z}^{signal}$ at 1064 nm does not appear absorbed over the 7$\mu$m length of the waveguide. 

\begin{figure}[htbp]
   \begin{minipage}[c]{.5\linewidth}
   \center
   \includegraphics[scale=0.675]{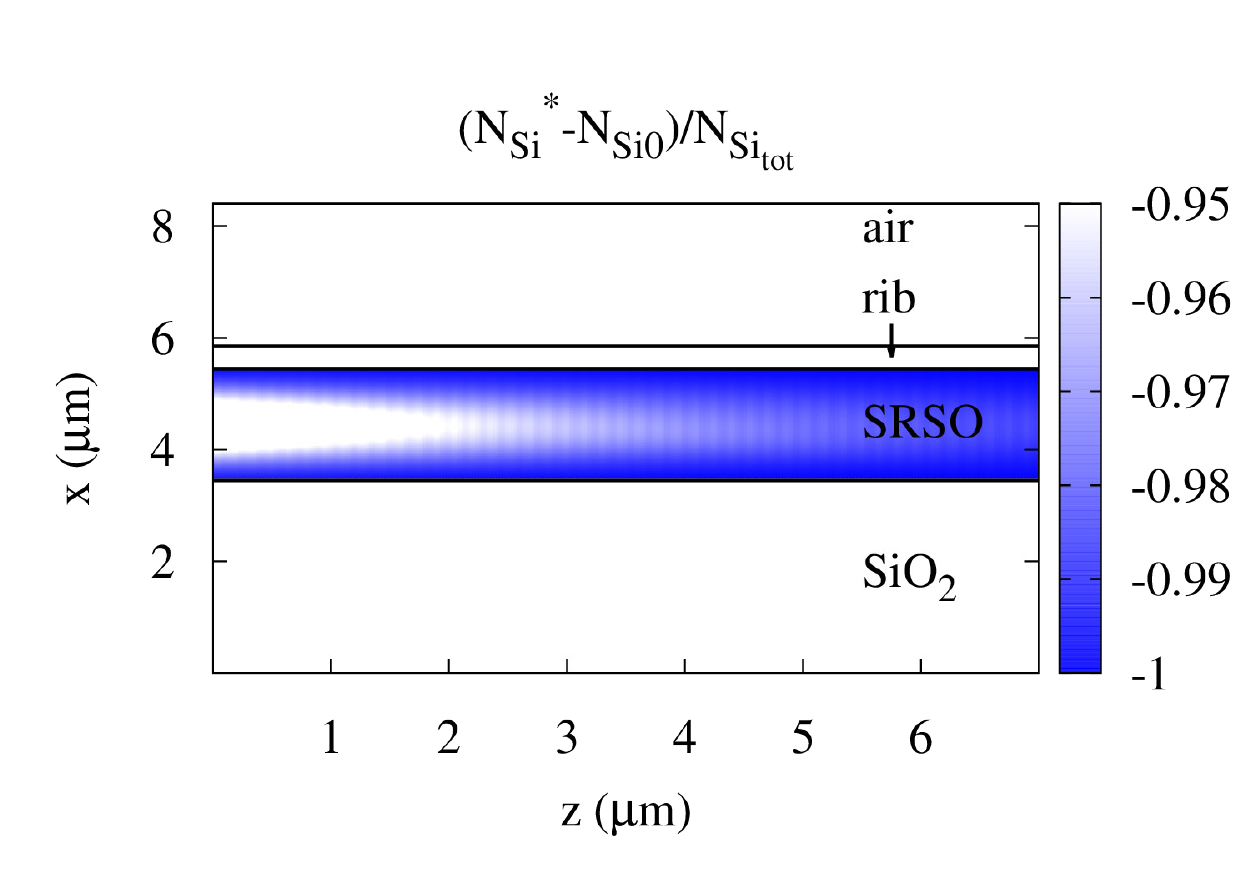}
   \end{minipage} \hfill
   \begin{minipage}[c]{.5\linewidth}
   \center
   \includegraphics[scale=0.675]{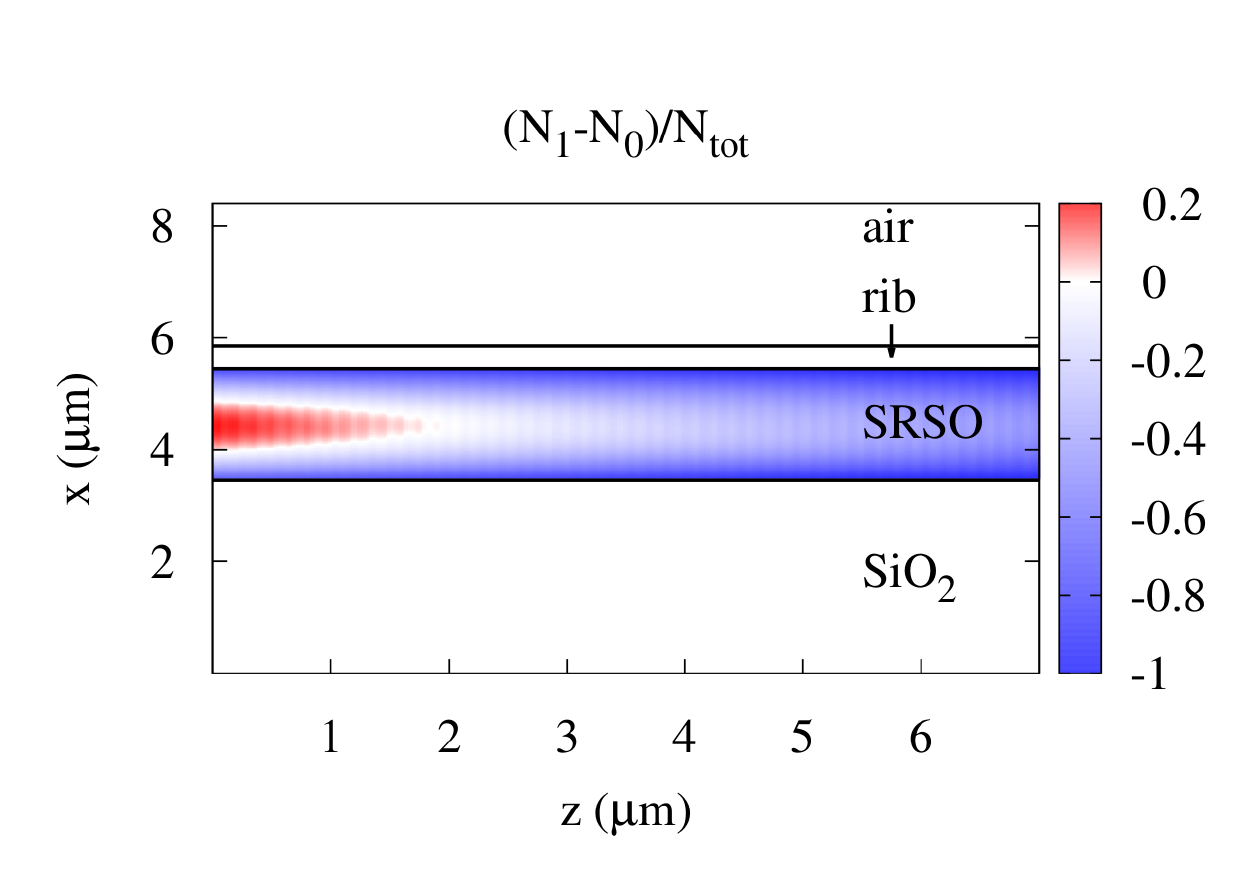}
   \end{minipage}
   \caption{Relative population differences along the direction of propagation of Si-ng (on the left) and for the erbium ions (on the right) in SiO$_x$:Si-ng:Er\up{3+} system for a pump power equal to 1000 mW.mm\up{-2}}
   \label{diffpopuEr}
\end{figure}

\begin{figure}[htbp]
   \begin{minipage}[c]{.5\linewidth}
   \center
   \includegraphics[scale=0.675]{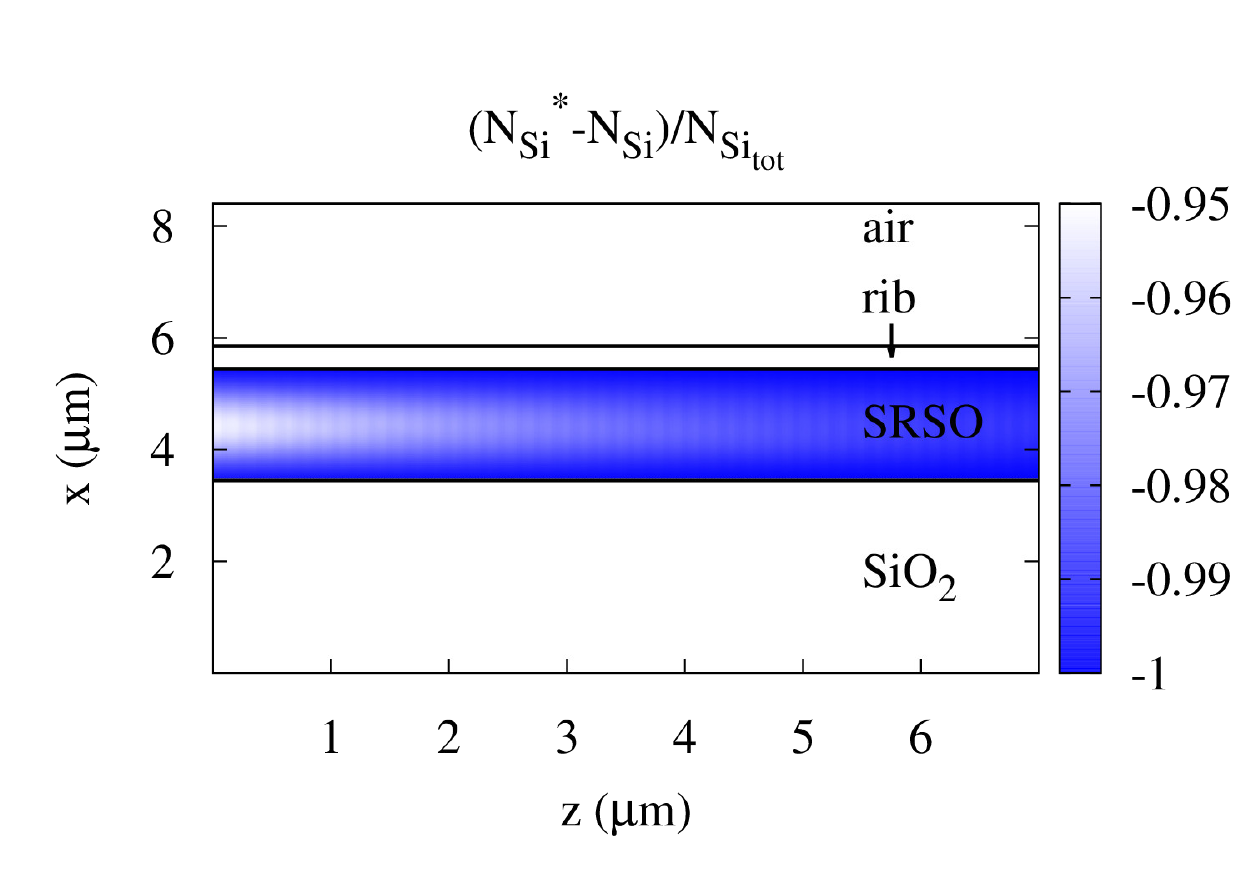}
   \end{minipage} \hfill
   \begin{minipage}[c]{.5\linewidth}
   \center
   \includegraphics[scale=0.675]{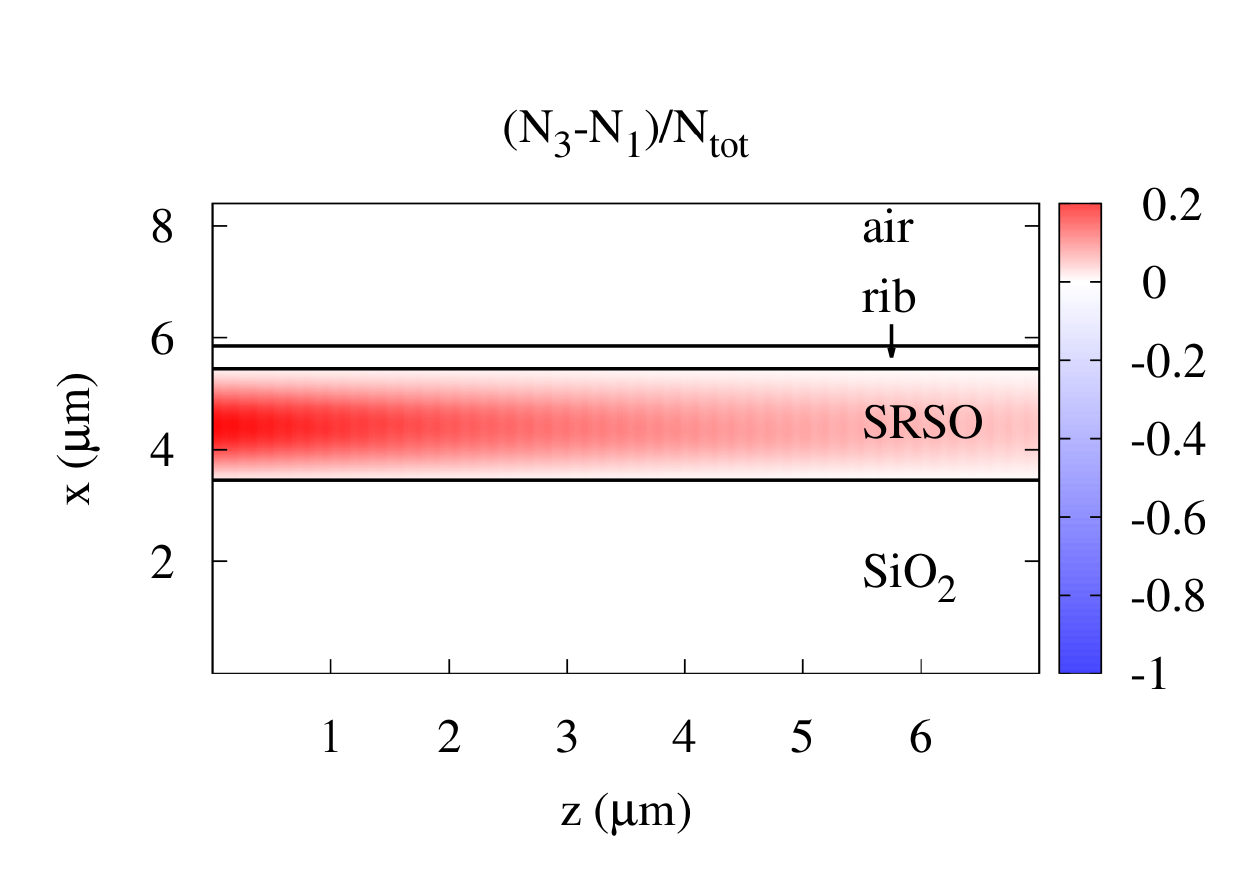}
   \end{minipage}
   \caption{Relative population differences along the direction of propagation of Si-ng (on the left) and for the neodymium ions (on the right) in SiO$_x$:Si-ng:Nd\up{3+} system for a pump power equal to 1000 mW.mm\up{-2}}
   \label{diffpopuNd}
\end{figure}

The two loops ADE-FDTD algorithm allows to compute tridimensional population distributions in different states. For a transition of interest, $N_{high}$ and $N_{low}$ represent respectively the higher and lower levels of the transition. We present the relative population difference (RPD) $\frac{N_{high}-N_{low}}{N_{tot}}$ distributions in the waveguide in a longitudinal section view along the propagation axis for the transition in Si-ng, and the 1532 nm transition of Er\up{3+} (figure \ref{diffpopuEr}) and the 1064 nm transition of Nd\up{3+} (figure \ref{diffpopuNd}) for a pump power equal to 10$^3$ mW.mm\up{-2}.

In case of Si-ng (figures \ref{diffpopuEr} left, \ref{diffpopuNd} left), the decreasing RPD profile with waveguide length is consistent with the pump profile \textbf{R}$_\mathrm{z}^{pump}$(z) decrease figure \ref{longi} due to the absorption of the pump field by the Si-ng. For erbium ions (figures \ref{diffpopuEr} right), the RPD of Nd\up{3+} ions shows a decrease with propagation length in the waveguide. This decrease is characteristic of the pump strong absorption due to the presence of the nanograins as shown in our previous paper \cite{fafin2013modeling}. The RPD of Er\up{3+} ions remains positive (inversion population is realized) over a length of 1.5 $\mu$m, beyond which it becomes negative witnessing the threshold effect occurring with three-levels system. 
For neodymium ions (figures \ref{diffpopuNd} right), the RPD of Nd\up{3+} ions shows also a decrease with propagation length in the waveguide which is also due to pump strong absorption by nanograins. The RPD of Nd\up{3+} ions remains positive (inversion population is realized) along the whole waveguide length witnessing the four-level system behavior, where the level 1 is depopulated quickly to the ground level leading to $N_3$ $\gg$ $N_1$.

\begin{figure}[htbp]
   \begin{minipage}[l]{.5\linewidth}
   \includegraphics[scale=0.33,trim=0mm 0mm 0mm 0mm,clip=true]{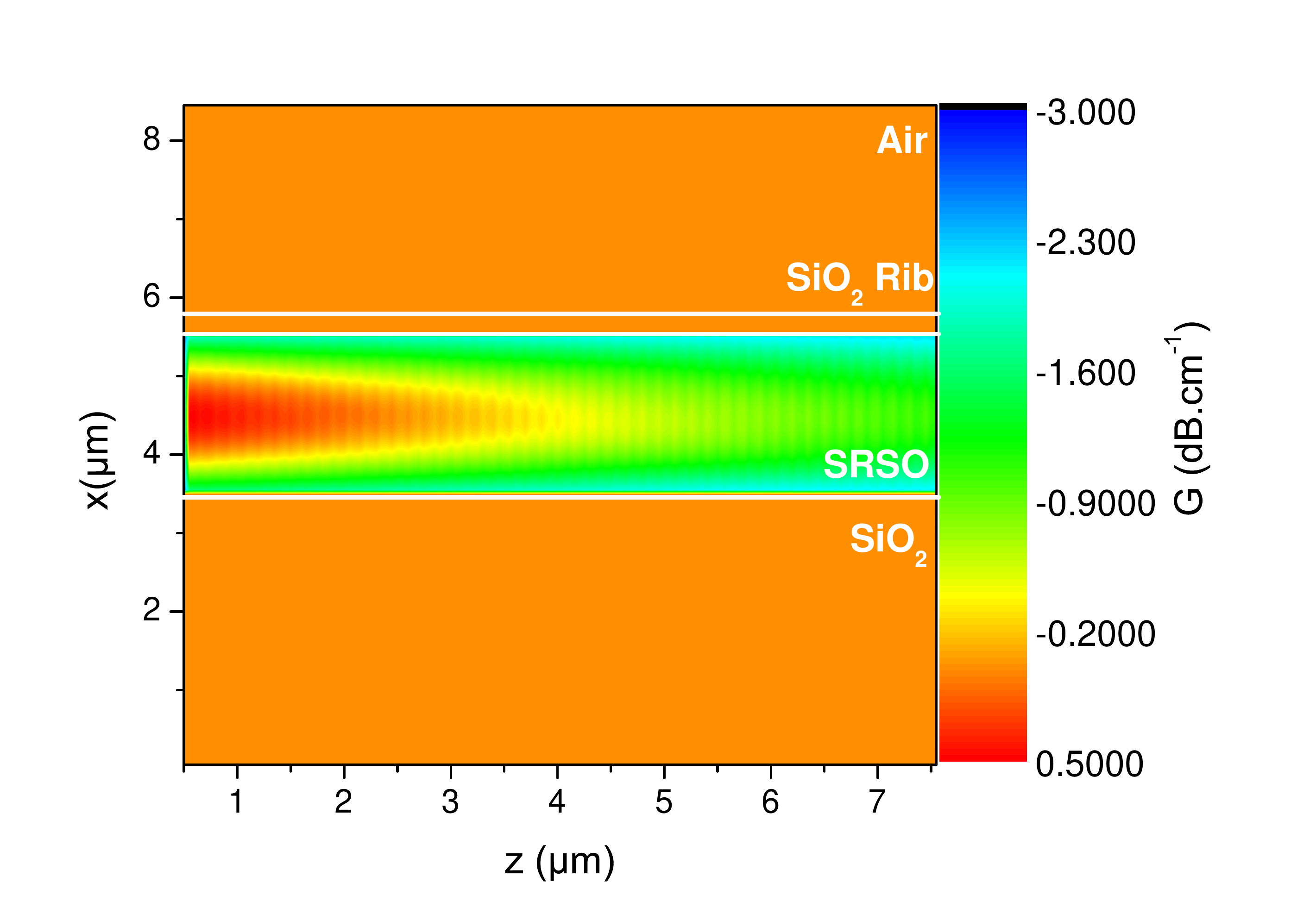}
   \end{minipage} \hfill
   \begin{minipage}[l]{.5\linewidth}
   \includegraphics[scale=0.33,trim=0mm 0mm 0mm 0mm,clip=true]{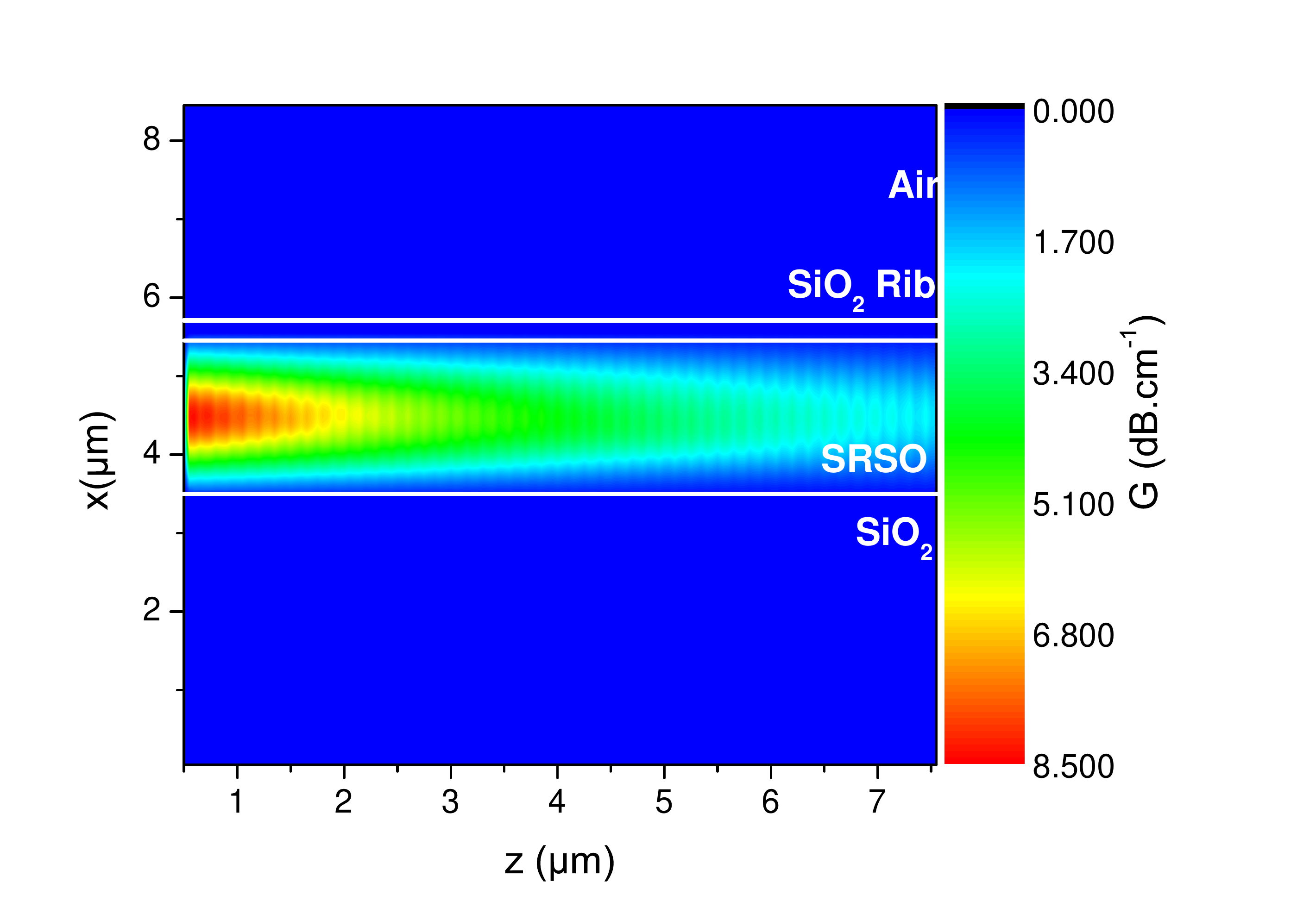}
   \end{minipage}
   \caption{Gross gain per unit length (dB.cm\up{-1}) along the direction of propagation in the case of erbium ions (on the left) and neodymium ions (on the right) for a pump power equal to 1000 mW.mm\up{-2}}
\label{cartesgain}
\end{figure}

\begin{figure}[htbp]
\centering
\scalebox{1.0}{\input{gain_nd_er.tex}}
\caption{Local gross gain per unit length as a function of the pumping power density for a waveguide doped with Nd\up{3+}(open circle) and a waveguide doped with Er\up{3+}(open square) recorded at x = 4.5 $\mu$m and y = 8.55 $\mu$m (center of the XY section of the active layer) and z = 0 (beginning of the waveguide). Losses found by Pirasteh et al \cite{Pirasteh2012} and by Navarro-Urrios et al \cite{navarro2011copropagating} are reported respectively in blue and red dashed line.}
\label{gain nd er}
\end{figure}
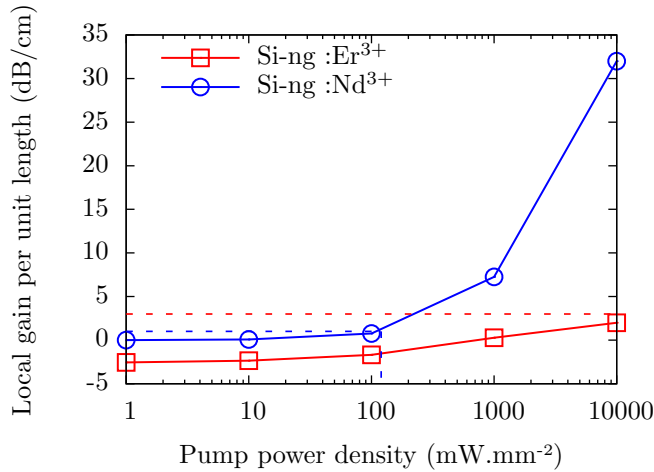

This co-propagating pumping configuration allows to study locally the waveguides with various pumping values in one calculation. Indeed from population levels distribution N$_i$(x,y,z) we can calculated the gross gain per unit length (dB.cm\up{-1}) at the signal wavelength by equation \ref{gain}.
\begin{equation}
g_{dB/cm}(x,y,z)=\frac{10}{ln10}\left( \sigma_{em}N_{high}(x,y,z) - \sigma_{abs}N_{low}(x,y,z)\right)
\label{gain}
\end{equation}
N$_{high}$ and N$_{low}$ are respectively the higher and lower levels of the considered transition and $\sigma_{abs}$ and $\sigma_{em}$ are respectively absorption and emission cross sections. For erbium ions, we use N$_{high}$ = N$_1$ and N$_{low}$=N$_0$ and for neodymium ions N$_{high}$ = N$_3$ and N$_{low}$ = N$_1$ (figure 6). Assuming equal emission and absorption cross sections, the gross gain per unit length longitudinal distribution for both rare earth ions are plotted in figure \ref{cartesgain}.
In case of erbium ions, figure \ref{cartesgain}a shows a positive (in red) gain per unit length only in the first 1.5 $\mu$m of the waveguide.  Elsewhere, the gain per unit length remains negative. For neodymium ions, the figure \ref{cartesgain}b shows a positive gain per unit length that remains everywhere positive in the waveguide. On these two dimensional distributions of the gross gain per unit length, the local gross gain per unit length can be extracted from anywhere in the active layer. We choose to take the local gross gain per unit length at the beginning of the waveguide in the center of the active layer (x = 4.5 $\mu$m and y = 8.55 $\mu$m) and in the first cell of the active layer of the waveguide where the pumping flux was recorded. We perform calculations with different pumping power densities at this extraction position and obtain the local gross gain per unit length plotted for different pumping power densities in figure \ref{gain nd er}.

For Er\up{3+} doped waveguide, above a threshold pump power of 1550 mW.mm\up{-2}, a positive gross gain is reached which increases up to 2 dB.cm\up{-1} for the highest pump power density simulated. In order to estimate the net gain, we must account for the background losses such as those found experimentally by Navarro-Urrios et al \cite{navarro2011copropagating}. They found on comparable waveguide losses of about 3 dB.cm\up{-1} at 1532 nm, making it impossible to reach a positive net gain in this range of pump power density. In case of Nd\up{3+} doped waveguide, we find that the optical gain remains positive over the whole power range and it increases up to 30 dB.cm\up{-1} for the highest pump power of $10^4$ mW.mm\up{-2}. Taking into account experimental background losses of 0.8 dB.cm\up{-1} found by Pirasteh et al \cite{Pirasteh2012} in a similar system, we can estimate a net gain per unit length. Figure \ref{gain nd er} shows that for a pump power above 130 mW.mm\up{-2} the losses (dashed line) can be compensated leading to a net gain. The gross gain per unit length obtained for Nd\up{3+} remains higher that the one obtained for Er\up{3+} whatever the pump power density range. The gross gain per unit length difference between Nd\up{3+} and Er\up{3+} doped waveguides increases  also with pump power density. Difference in Si-ng/RE transfer efficiency linked to the levels dynamics as well as to the difference in absorption/emission cross sections\cite{fafin2014theoretical} ($\sigma_{Er}$ =$6.10^{21}$ $cm^{2}$ against $\sigma_{Nd}$ =$1.10^{19}$ $cm^{2}$) may explain this gross gain per unit length feature. The low gain and a short length of positive population inversion obtained by modeling of Er\up{3+} doped based waveguide on the broad range of pump power make impossible the achievement of an optical amplifier with this configuration which would compete with commercially available systems. The gross gain per unit length reachable with Nd\up{3+}, about one order of magnitude larger than the one obtained with Er\up{3+}, could lead to a significant amplification. Nowadays, there is no commercially available comparable optical amplifier based on Nd\up{3+} emission bands. However, Nd\up{3+} doped Aluminum oxide channel waveguide amplifiers developed by Yang et al\cite{yang2010high} show a maximal internal gain of 6 dB.cm\up{-1} for a pump power of 45 mW.

%

\section{Conclusion}

A new two loops ADE-FDTD algorithm has been developed that allows to describe the spatial distribution of the electromagnetic field and steady state population levels in an active optical waveguide. The multi-scale times issue of such a system consisting in crippling calculation duration due to large difference between FDTD time step and population level lifetimes has been overcome by this algorithm. Moreover, we proposed a calibration method of the  $i\rightarrow j$ transition linewidth $\Delta\omega _{ij}$ according to the experimental cross section, making a possible comparison between experimental and theoretical studies. We apply our algorithm to a strip loaded waveguide whose active layer is constituted of a silicon rich silicon oxide layer containing silicon nanograins and doped with either Er\up{3+} or with Nd\up{3+} ions. Our method permits to determine a three dimensional distribution of gross gain per unit length in the waveguide in steady state. We have demonstrated that, to obtain a net positive gain per unit length in SiO$_x$:Si-ng:RE\up{3+} based waveguide, the neodymium ions are more suitable than the erbium ions. For the latter, the theoretical maximum gross gain per unit length of 2 dB.cm\up{-1} at 1532 nm ($10^4$ mW.mm\up{-2}) does not compensate background losses experimentally estimated to 3 dB.cm\up{-1}. On the contrary, the use of neodymium ions leads to a gross gain per unit length of 30 dB.cm\up{-1} at 1064 nm ($10^4$ mW.mm\up{-2}). Moreover the background losses are compensated above a pump power threshold of 130 mW.mm\up{-2}. This theoretical demonstration of a large gross gain per unit length for a Nd\up{3+} doped active layer may justify further experimental work in order to achieve Nd\up{3+} doped silicon based waveguide optical amplifier or laser. Further studies may be performed exploring other concentrations of rare earth ions and Si-ng, other rare earth ions and other pumping configurations in order to investigate the possibility of achieving larger gain. This method may be applied successfully to describe steady states of other kind of emitters (quantum dots, quantum wells, Dyes...) and in other configuration (VCELs, down-converting layers...).

\acknowledgments     
The authors are grateful to the French National Research Agency, which supported this work through the Nanoscience and Nanotechnology program (DAPHNES ANR-08-NANO-005 project), to the project EMC3 Labex ASAP and to the Centre de ressources informatiques de Haute-Normandie (CRIHAN) for computing facilities.

\bibliography{theseAF}   
\bibliographystyle{spiebib}   

\end{document}

%% file: SRSO_Er2.tex
        \begin{tikzpicture}
          \node at (0,0){\includegraphics[scale=1.0]{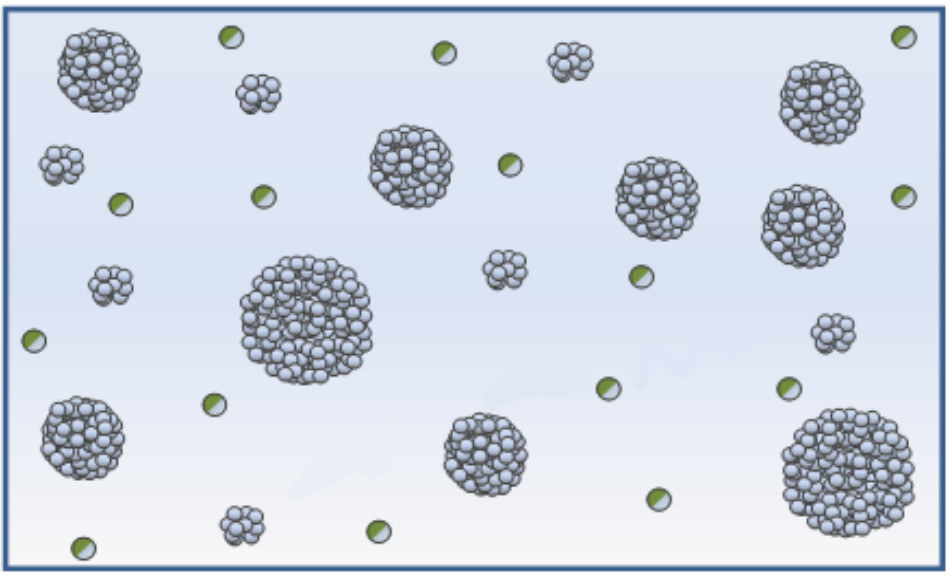}};
          \draw [thick,Blue,decorate,->,>=latex,decoration={snake,post length=2mm,segment length=5pt,amplitude=2pt}] (0.85,-0.575) -- (1.4,-0.1);
          \draw [ultra thick,SaddleBrown,->] (0.1,-0.8) to [bend left=55] (0.7,-0.5);
          \draw [thick,Black,decorate,->,>=latex,decoration={snake,post length=2mm,segment length=5pt,amplitude=2pt}] (-0.2,-1) -- (-1,-1.1);
          \draw [->,>=latex,black] (-1.1,0.75) node [left] {Si-ng} -- (-0.65,0.75);
          \draw [<-,>=latex,black] (0.275,0.775) -- (0.75,0.9);
          \draw (0.75,0.75) node [above right] {Er\up{3+}};
          \draw (2.2,1.1) node [above right] {\textbf{(a)}};
        \end{tikzpicture} 

%% file: SRSO_Nd2.tex
        \begin{tikzpicture}
          \node at (0,0){\includegraphics[scale=1.0]{SRSO_Nd2.jpeg}};
          \draw [thick,Red,decorate,->,>=latex,decoration={snake,post length=2mm,segment length=5pt,amplitude=2pt}] (0.85,-0.575) -- (1.4,-0.1);
          \draw [thick,Blue,decorate,->,>=latex,decoration={snake,post length=2mm,segment length=5pt,amplitude=2pt}] (0.85,-0.575) -- (1.4,-0.9);
          \draw [thick,Green,decorate,->,>=latex,decoration={snake,post length=2mm,segment length=5pt,amplitude=2pt}] (0.85,-0.575) -- (1.6,-0.57);
          \draw [ultra thick,SaddleBrown,->] (0.1,-0.8) to [bend left=55] (0.7,-0.5);
          \draw [thick,Black,decorate,->,>=latex,decoration={snake,post length=2mm,segment length=5pt,amplitude=2pt}] (-0.2,-1) -- (-1,-1.1);
          \draw [->,>=latex,black] (-1.1,0.75) node [left] {Si-ng} -- (-0.65,0.75);
          \draw [<-,>=latex,black] (0.275,0.775) -- (0.75,0.9);
          \draw (0.75,0.75) node [above right] {Nd\up{3+}};
          \draw (2.2,1.1) node [above right] {\textbf{(b)}};
        \end{tikzpicture} 

%% file: algoclassique.tex
\def\ech{1.2}
\begin{tikzpicture}
    	\tikzstyle{fond}=[thick,rounded corners=4pt, fill=yellow!20]
        \tikzstyle{fond2}=[thick,rounded corners=4pt, fill=green!10]
        \tikzstyle{rect}=[rectangle,draw,fill=red!20,text width=3cm,text centered]
 		\tikzstyle{losange}=[diamond,draw,aspect=1.5,thick,fill=blue!20,scale=\ech]
        \tikzstyle{fleche}=[->,thick,>=latex,rounded corners=4pt]

        \draw [fond] (-3,-2) rectangle (3,6);
      \node[rect] (EH) at (0,5) {$\mathbf{E},\mathbf{H}$ \textcolor{gray}{(FDTD)}};
      \node[rect] (P) at (0,4) {$\mathbf{P}_{ij}$ \textcolor{gray}{(ADE)}};
    \node[rect] (N) at (0,3) {$N_i$};
    \node[losange] (test) at (0,1) {$\frac{dN_i}{dt}=0?$};
    \node[rect] (end) at (0,-1) {END};

    \draw[fleche] (EH) -- (P);
    \draw[fleche] (P)-- (N);
    \draw[fleche] (test)-l (-2.5,1) -l (-2.5,5) -- (EH);
    \draw[fleche] (N)--(test);
    \draw[fleche] (test) -- (end);

    \draw (test.west) node{$ \bullet $} node[below]{no};
    \draw (test.south) node{$ \bullet $} node[below right]{yes};
    \draw (2.2,5.25) node [above right] {\textbf{(a)}};
  \end{tikzpicture} 

%% file: nouvelalgo.tex
\def\ech{0.7}
      \begin{tikzpicture}
        \tikzstyle{fond}=[thick,rounded corners=4pt, fill=yellow!20]
        \tikzstyle{fond2}=[thick,rounded corners=4pt, fill=green!10]
        \tikzstyle{rect}=[rectangle,draw,fill=red!20,text width=3cm,text centered]
		\tikzstyle{losange}=[diamond,draw,aspect=1.5,thick,fill=blue!20,scale=\ech]
        \tikzstyle{fleche}=[->,thick,>=latex,rounded corners=4pt]

          \draw [fond] (-3,0) rectangle (2.5,6);
          \draw [fond2] (3,0) rectangle (8.5,6);
          \node[rect] (EH) at (0,5) {$\mathbf{E},\mathbf{H}$ \textcolor{gray}{(FDTD)}};
          \node[rect] (P) at (0,4)  {$\mathbf{P}_{ij}$ \textcolor{gray}{(ADE)}};
          \node[losange] (EdP) at (0,2) {$\left<\mathbf{E} \frac{d\mathbf{P}_{ij}}{dt}\right>(t)=cst?$};

        \node[rect] (N) at (6,5) {$N_i$ (steady state)};
        \node[losange] (testN) at (6,3) {$N_i^R-N_i^{R-1}=cst?$};
        \node[rect] (fin) at (6,1) {END};

        \draw[fleche] (EH) -- (P);
        \draw[fleche] (P)-- (EdP);
        \draw[fleche] (EdP) -l (-2.5,2) -l (-2.5,5) -- (EH);
        \draw[fleche] (EdP) -l (3.5,2) -l (3.5,5) -- (N);

        \draw[fleche] (N) -- (testN);
        \draw[fleche] (testN) -- (fin);
        \draw[thick] (testN.west) -- (3.6,3) arc(0:180:0.1);
        \draw[fleche] (3.4,3) -l (2,3) -l (2,5) -- (EH.east);

        \draw (EdP.west) node{$ \bullet $} node[below]{no};
        \draw (EdP.east) node{$ \bullet $} node[below]{yes};
        \draw (testN.west) node{$ \bullet $} node [below] {no};
        \draw (testN.south) node{$\bullet$} node[below right] {yes};
        \draw (1.7,5.25) node [above right] {\textbf{(b)}};
    \end{tikzpicture}

%% file: schema_erbium_complet.tex
  \begin{tikzpicture}
    \tikzset{flechenr/.style={->,>=latex,decorate, decoration={segment length=5pt,snake,post length=1mm}}}

    \draw [fill,blue!20] (0,-0.25) rectangle (6,0.25);
    \draw [fill,blue!40] (1,-0.25) rectangle (2,1);
    
    \draw [fill,blue!20] (0,5) rectangle (6,5.5);
    \draw [fill,blue!40] (1,5.5) rectangle (2,4.5);
    
    \draw (6,5.25) node [left] {SiO$_2$};
    \draw (6,0) node [left] {SiO$_2$};
    
    \draw (1.5,0) node {Si-ng};

    \draw [red!10,fill=red!10] (3.9,0.55) rectangle (6.05,4.7);

    
    \draw (1,1) node [left] {Si$ $};
    \draw (1,4.5) node [left] {Si$^*$};
    \draw [black,fill=white] (1.3,1) circle (0.15); 
    \draw [black,fill=gray] (1.3,5.2) circle (0.15);
    \draw [black,fill=black] (1.8,4.5) circle (0.15);
    
    \draw [very thick,<->,>=latex,DarkGreen] (1.3,1.15) -- (1.3,5.05);
    \draw [flechenr] (1.4,5.05) -- (1.7,4.5);
    
    \draw [thick,DarkRed,decorate,->,>=latex,decoration={snake,post length=3mm}] (-1,2) -- (1.1,1.05);
    \draw [DarkRed] (-1,2.2) node [right]{$\lambda=488$ nm};
    
    \draw [thick,red] (4,4.5) -- (6,4.5) node [black,right] {3 (\up{4}I$_{9/2}$)};
    \draw [thick,red] (4,4) -- (6,4) node [black,right] {2 (\up{4}I$_{11/2}$)};
    \draw [thick,red] (4,3.5) node [black,left] {1532 nm} -- (6,3.5) node [black,right] {1 (\up{4}I$_{13/2}$)};
    \draw [thick,red] (4,1)-- (6,1) node [black,right] {0 (\up{4}I$_{15/2}$)};
    \draw [red] (5,0.5) node [above] {Er\up{3+}};
    
    \draw [flechenr] (5.2,4.5) -- (5.9,4);
    \draw [flechenr] (5.2,4) -- (5.9,3.5);
    \draw [flechenr] (5,4) -- (5.9,1);
    
    \draw [->,>=latex,very thick, red] (4.2,3.5) -- (4.2,1);
    
    \draw [->,>=latex, very thick, MediumVioletRed] (4.5,3.5) -- (4.5,4.5);
    \draw [fill,MediumVioletRed] (4.5,3.5) circle (0.1);
    \draw [->,>=latex, very thick, MediumVioletRed] (4.8,3.5) -- (4.8,1);
    \draw [fill,MediumVioletRed] (4.8,3.5) circle (0.1);
    
    \draw [SaddleBrown,very thick, ->, >=latex] (2,4.5) to[bend left] (4,4.5);
    \draw [SaddleBrown] (3,4.7) node [text width = 2 cm,text centered,below] {Energy Transfer};
    \draw (0.05,4.85) node [above right] {\textbf{(a)}};
  \end{tikzpicture}

%% file: schema_neodyme_complet.tex
  \begin{tikzpicture}
    \tikzset{flechenr/.style={->,>=latex,decorate, decoration={segment length=5pt,snake,post length=1mm}}}

    \draw [fill,blue!20] (0,-0.25) rectangle (6,0.25);
    \draw [fill,blue!40] (1,-0.25) rectangle (2,1);
    
    \draw [fill,blue!20] (0,5) rectangle (6,5.5);
    \draw [fill,blue!40] (1,5.5) rectangle (2,4.5);
    
    \draw (6,5.25) node [left] {SiO$_2$};
    \draw (6,0) node [left] {SiO$_2$};
    
    \draw (1.5,0) node {Si-ng};

    \draw [red!10,fill=red!10] (3.9,0.55) rectangle (6.05,4.7);

    
    \draw (1,1) node [left] {Si$ $};
    \draw (1,4.5) node [left] {Si$^*$};
    \draw [black,fill=white] (1.3,1) circle (0.15); 
    \draw [black,fill=gray] (1.3,5.2) circle (0.15);
    \draw [black,fill=black] (1.8,4.5) circle (0.15);
    
    \draw [very thick,<->,>=latex,DarkGreen] (1.3,1.15) -- (1.3,5.05);
    \draw [flechenr] (1.4,5.05) -- (1.7,4.5);
    
    \draw [thick,DarkRed,decorate,->,>=latex,decoration={snake,post length=3mm}] (-1,2) -- (1.1,1.05);
    \draw [DarkRed] (-1,2.2) node [right]{$\lambda=488$ nm};
    
    \draw [thick,red] (4,4.5) -- (6,4.5) node [black,right] {4 (\up{4}F$_{5/2}$+\up{2}H$_{9/2}$)};
    \draw [thick,red] (4,4) -- (6,4) node [black,right] {3 (\up{4}F$_{3/2}$)};
    \draw [thick,red] (4,2) node [black,left] {1340 nm} -- (6,2) node [black,right] {2 (\up{4}I$_{13/2}$)};
    \draw [thick,red] (4,1.5) node [black,left] {1064 nm} -- (6,1.5) node [black,right] {1 (\up{4}I$_{11/2}$)};
    \draw [thick,red] (4,1) node [black,left] {945 nm} -- (6,1) node [black,right] {0 (\up{4}I$_{9/2}$)};
    \draw [red] (5,0.5) node [above] {Nd\up{3+}};
    
    \draw [flechenr] (5.2,4.5) -- (5.9,4);
    \draw [flechenr] (5.2,2) -- (5.9,1.5);
    \draw [flechenr] (5.2,1.5) -- (5.9,1);
    
    \draw [->,>=latex,very thick, gray] (4.2,4) -- (4.2,1);
    \draw [->,>=latex,very thick, red] (4.5,4) -- (4.5,1.5);
    \draw [->,>=latex,very thick, gray] (4.8,4) -- (4.8,2);
    
    \draw [SaddleBrown,very thick, ->, >=latex] (2,4.5) to[bend left] (4,4.5);
    \draw [SaddleBrown] (3,4.7) node [text width = 2 cm,text centered,below] {Energy Transfer};
    \draw (0.05,4.85) node [above right] {\textbf{(b)}};
  \end{tikzpicture}

%% file: gain_nd_er.tex
\begingroup
  \makeatletter
  \providecommand\color[2][]{%
    \GenericError{(gnuplot) \space\space\space\@spaces}{%
      Package color not loaded in conjunction with
      terminal option `colourtext'%
    }{See the gnuplot documentation for explanation.%
    }{Either use 'blacktext' in gnuplot or load the package
      color.sty in LaTeX.}%
    \renewcommand\color[2][]{}%
  }%
  \providecommand\includegraphics[2][]{%
    \GenericError{(gnuplot) \space\space\space\@spaces}{%
      Package graphicx or graphics not loaded%
    }{See the gnuplot documentation for explanation.%
    }{The gnuplot epslatex terminal needs graphicx.sty or graphics.sty.}%
    \renewcommand\includegraphics[2][]{}%
  }%
  \providecommand\rotatebox[2]{#2}%
  \@ifundefined{ifGPcolor}{%
    \newif\ifGPcolor
    \GPcolortrue
  }{}%
  \@ifundefined{ifGPblacktext}{%
    \newif\ifGPblacktext
    \GPblacktexttrue
  }{}%
  \let\gplgaddtomacro\g@addto@macro
  \gdef\gplbacktext{}%
  \gdef\gplfronttext{}%
  \makeatother
  \ifGPblacktext
    \def\colorrgb#1{}%
    \def\colorgray#1{}%
  \else
    \ifGPcolor
      \def\colorrgb#1{\color[rgb]{#1}}%
      \def\colorgray#1{\color[gray]{#1}}%
      \expandafter\def\csname LTw\endcsname{\color{white}}%
      \expandafter\def\csname LTb\endcsname{\color{black}}%
      \expandafter\def\csname LTa\endcsname{\color{black}}%
      \expandafter\def\csname LT0\endcsname{\color[rgb]{1,0,0}}%
      \expandafter\def\csname LT1\endcsname{\color[rgb]{0,1,0}}%
      \expandafter\def\csname LT2\endcsname{\color[rgb]{0,0,1}}%
      \expandafter\def\csname LT3\endcsname{\color[rgb]{1,0,1}}%
      \expandafter\def\csname LT4\endcsname{\color[rgb]{0,1,1}}%
      \expandafter\def\csname LT5\endcsname{\color[rgb]{1,1,0}}%
      \expandafter\def\csname LT6\endcsname{\color[rgb]{0,0,0}}%
      \expandafter\def\csname LT7\endcsname{\color[rgb]{1,0.3,0}}%
      \expandafter\def\csname LT8\endcsname{\color[rgb]{0.5,0.5,0.5}}%
    \else
      \def\colorrgb#1{\color{black}}%
      \def\colorgray#1{\color[gray]{#1}}%
      \expandafter\def\csname LTw\endcsname{\color{white}}%
      \expandafter\def\csname LTb\endcsname{\color{black}}%
      \expandafter\def\csname LTa\endcsname{\color{black}}%
      \expandafter\def\csname LT0\endcsname{\color{black}}%
      \expandafter\def\csname LT1\endcsname{\color{black}}%
      \expandafter\def\csname LT2\endcsname{\color{black}}%
      \expandafter\def\csname LT3\endcsname{\color{black}}%
      \expandafter\def\csname LT4\endcsname{\color{black}}%
      \expandafter\def\csname LT5\endcsname{\color{black}}%
      \expandafter\def\csname LT6\endcsname{\color{black}}%
      \expandafter\def\csname LT7\endcsname{\color{black}}%
      \expandafter\def\csname LT8\endcsname{\color{black}}%
    \fi
  \fi
  \setlength{\unitlength}{0.0500bp}%
  \begin{picture}(5040.00,3600.00)%
    \gplgaddtomacro\gplbacktext{%
      \csname LTb\endcsname%
      \put(814,704){\makebox(0,0)[r]{\strut{}-5}}%
      \put(814,1033){\makebox(0,0)[r]{\strut{} 0}}%
      \put(814,1362){\makebox(0,0)[r]{\strut{} 5}}%
      \put(814,1691){\makebox(0,0)[r]{\strut{} 10}}%
      \put(814,2020){\makebox(0,0)[r]{\strut{} 15}}%
      \put(814,2348){\makebox(0,0)[r]{\strut{} 20}}%
      \put(814,2677){\makebox(0,0)[r]{\strut{} 25}}%
      \put(814,3006){\makebox(0,0)[r]{\strut{} 30}}%
      \put(814,3335){\makebox(0,0)[r]{\strut{} 35}}%
      \put(946,484){\makebox(0,0){\strut{} 1}}%
      \put(1870,484){\makebox(0,0){\strut{} 10}}%
      \put(2795,484){\makebox(0,0){\strut{} 100}}%
      \put(3719,484){\makebox(0,0){\strut{} 1000}}%
      \put(4643,484){\makebox(0,0){\strut{} 10000}}%
      \put(176,2019){\rotatebox{-270}{\makebox(0,0){\strut{}Local gain per unit length (dB/cm)}}}%
      \put(2794,154){\makebox(0,0){\strut{}Pump power density (mW.mm\up{-2})}}%
    }%
    \gplgaddtomacro\gplfronttext{%
      \csname LTb\endcsname%
      \put(1933,3162){\makebox(0,0)[l]{\strut{}Si-ng:$\mathrm{Er^{3+}}$}}%
      \csname LTb\endcsname%
      \put(1933,2942){\makebox(0,0)[l]{\strut{}Si-ng:$\mathrm{Nd^{3+}}$}}%
    }%
    \gplbacktext
    \put(0,0){\includegraphics{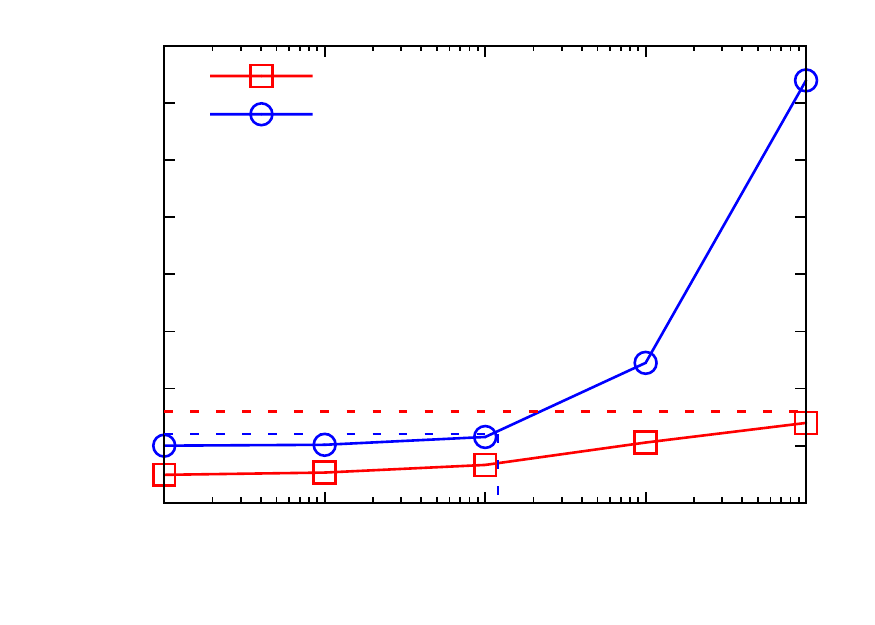}}%
    \gplfronttext
  \end{picture}%
\endgroup